\documentclass[a4paper,11pt]{article}
\usepackage{jheppub} 
\usepackage{lineno}
\usepackage{caption}
\usepackage{subcaption}
\usepackage{physics}
\usepackage{tikz-feynman}

\title{\boldmath Coherent $\gamma^*$-nucleus scattering and coherent nuclear states}

\author{A. H. Mueller}
\affiliation{Department of Physics\\
Columbia University\\
New York, NY 10027}

\emailAdd{ahm4@columbia.edu}

\abstract{
In the context of a McLerran-Venugopalan (MV) model for a large nucleus, coherent scattering of a virtual photon on that nucleus is evaluated in the $A_-=0$ gauge, the gauge appropriate for the target nucleus. The evaluation of the scattering in $A_-=0$ gauge is very intricate compared to the usual $A_+=0$ gauge evaluation natural for the scattering process, but has the advantage of directly giving the scattering in terms of a partonic description of the nucleus. In the limit where a tagged forward jet puts the dipole-nucleus scattering in the saturation regime the coherent reactions are equal to the inelastic reactions. In terms of the nuclear wave function the coherent reactions come from color singlet and zero total transverse momentum quark-antiquark pairs in the wave function and in the saturation regime the nuclear wave function is a coherent state for these pairs. In the saturation region half of all quarks (or antiquarks) come from zero momentum and color charge pairs.
}

\begin{document}
\maketitle
\flushbottom

\section{Introduction}
\label{sec:intro}

This paper is focused on saturation of parton densities \cite{1} in the light cone wave function of a hadron or nucleus. The discussion will primarily concern quark saturation as it most easily shows up in electron scattering, however gluon saturation displays all the properties we shall discuss as well as having a much higher occupancy.\cite{2} (The analogs of \ref{eq:63} and \ref{eq:70} for gluon distributions would have an extra $\frac{1}{\alpha N_C}$ on the right hand side of these equations.) A main part of our focus is the comparison of how saturation appears in coherent nuclear reactions versus inelastic reactions. \cite{3} As we go along it should become apparent that parton saturation and attaining the unitarity limits in high energy dipole hadron scattering are dual versions of the same phenomenon. This is not new, however, the way in which we do our calculations makes this duality manifest. The calculation of black disc scattering of the dipole and the calculation of the high $(q\bar{q})$ density are identical.

Our target will be a McLerran-Venugopalan (MV) nucleus \cite{4,5,6,7,8} on which we shall scatter a highly virtual transverse photon. In Sec.~\ref{sec:2} we take the nucleus at rest with the virtual photon having a large value of $q_+$ and where $2q_+/Q^2$ is also large. In this section we use $A_+=0$ light cone gauge the gauge most natural for this high energy scattering, however, a gauge not well suited for extracting parton distributions. We review the calculations of forward jet production in coherent $\gamma^*$-nucleus scattering, and how the single jet cross section is nicely expressed in terms of quark-antiquark dipole scattering on the nucleus. \cite{2} Then we turn to inelastic electron scattering where we focus on the (well known) total inelastic cross sections (DIS). In each of these cases, coherent and inelastic scattering, the cross sections are written in a way where the unitarity limit of strong scattering is clearly visible. However, although the cross sections for coherent and inelastic scattering are easily obtained, partonic interpretations are not easy to see in these $A_+=0$ gauge calculations.

In Sec.~\ref{sec:3} we redo the calculations of Sec.~\ref{sec:2}, but now using $A_-=0$ gauge. We begin by keeping the same frame of reference as in Sec.~\ref{sec:2} so that we imagine $x_+$ as the time variable but we use $A_-=0$ gauge. This cannot be done in light cone perturbation theory but is possible, if not straightforward, using Feynman graphs. This is an unusual gauge for the high energy scattering of a quark-antiquark dipole, having large $q_+$, on a target at rest. The benefit of doing the high energy scattering in this unusual way is that the calculation, once done, can be converted easily to a light cone perturbation theory calculation, now viewing the $\gamma^*$ as a probe of the target having $p_-$ large, and thus a partonic interpretation becomes available. \cite{2,9}

We begin Sec.~\ref{sec:3} with a momentum space calculation of a single coherent scattering of the $\gamma^*$ on the MV nucleus. Once this calculation is done we convert the result to a coordinate space dipole target scattering in order to compare with our calculations of Sec.~\ref{sec:2}. However, the multiple scattering problem will only be possible in coordinate space so we then, in Sec.~\ref{sec:3.2}, redo the single scattering case but now completely in coordinate space. Once this calculation is done the coherent multiple scattering situation is easily handled in Sec.~\ref{sec:3.3}, where we reproduce the results from Sec.~\ref{sec:2}. But now that the calculation has been done in $A_-=0$ gauge the partonic interpretation is straightforward.

Next, in Sec.~\ref{sec:3.4} we do single, inelastic scattering and in Sec.~\ref{sec:3.5} inelastic reactions with multiple scattering. These calculations agree with those of Sec.~\ref{sec:2} and are the final technical calculations necessary for a partonic interpretation, the topic to which we next turn.

In Sec.~\ref{sec:4} we note that the $A_-=0$ gauge scatterings that have just been done can be viewed in terms of partonic evolution of the target nucleus and the scattering calculations can be taken over to give the quark distribution of the target without further calculation. More precisely the scattering graphs of Fig.~\ref{fig:11} are identical to the partonic graphs of Fig.~\ref{fig:14}. We then use the scattering calculations of Sec.~\ref{sec:2} to obtain the saturated quark, and antiquark, distributions for elastic and inelastic reactions in \eqref{eq:63} and \eqref{eq:70}.

In Sec.~\ref{sec:5} we discuss the phenomenon of saturation. The main new result is that coherent elastic $\gamma^*$-nucleus scattering corresponds to the nucleus having zero total transverse momentum $q\bar{q}$-pairs which, when the individual quark momenta are below the saturation momentum, are arranged as a coherent state of such pairs. That is if one takes the light cone wave function of the nucleus and eliminates a zero total transverse momentum $q\bar{q}$ pair from the wave function one does not change the wave function, which in more physical terms is the same as saying that a $\gamma^*$ scattering on the nuclear wave function can produce a zero transverse momentum pair and leave the rest of the wave function still in its ground state. Thus the duality between reaching the unitarity limit in coherent elastic scattering is to have the nuclear wave function, in the saturation region, be a coherent state of $q\bar{q}$-pairs.

Finally we note that in the saturation regime one-half the quarks, and gluons, are in a coherent state and correspond to coherent reactions. Thus for example if one produces a $\mu$-pair in a nucleus-nucleus collision, with the transverse momentum of the pair in the saturation region, and the mass not so large as to make Sudakov corrections large, then the rate for the production of such $\mu$-pairs is four times as large as that predicted by the quark distributions coming only from inelastic electron-nucleus scattering.

\section{Projectile picture of coherent and inelastic scattering in $A_+=0$ gauge}
\label{sec:2}

\subsection{Coherent scattering}
\label{sec:2.1}

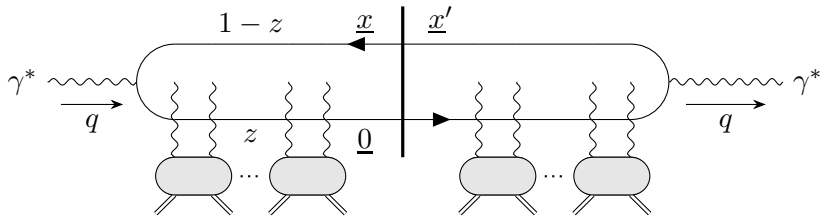
\begin{figure}[htbp]
\centering
\begin{tikzpicture}
\begin{feynman} 
\vertex (p1) {\(\gamma^*\)};
\vertex [right=of p1] (l);
\vertex [below right=0.707cm of l] (b1);
\vertex [right=0.5cm of b1] (b2);
\vertex [right=1.0cm of b2] (b3);
\vertex [right=0.5cm of b3] (b4);
\vertex [right=1.0cm of b4] (b5);
\vertex [right=1.0cm of b5] (b6);
\vertex [right=0.5cm of b6] (b7);
\vertex [right=1.0cm of b7] (b8);
\vertex [right=0.5cm of b8] (b9);
\vertex [above right=0.707cm of b9] (r);
\vertex [right=of r] (p2) {\(\gamma^*\)};
\vertex [above right=0.707cm of l] (t1);
\vertex [right=0.5cm of t1] (t2);
\vertex [right=1.0cm of t2] (t3);
\vertex [right=0.5cm of t3] (t4);
\vertex [right=1.0cm of t4] (t5);
\vertex [right=1.0cm of t5] (t6);
\vertex [right=0.5cm of t6] (t7);
\vertex [right=1.0cm of t7] (t8);
\vertex [right=0.5cm of t8] (t9);
\vertex [right=0.5cm of l] (m1);
\vertex [right=0.5cm of m1] (m2);
\vertex [right=1.0cm of m2] (m3);
\vertex [right=0.5cm of m3] (m4);
\vertex [right=1.0cm of m4] (m5);
\vertex [right=1.0cm of m5] (m6);
\vertex [right=0.5cm of m6] (m7);
\vertex [right=1.0cm of m7] (m8);
\vertex [right=0.5cm of m8] (m9);
\vertex [above=0.5cm of t5] (tt);
\vertex [below=0.5cm of b5] (bb);
\vertex [below=0.5cm of b1] (n1tl);
\vertex [below=0.5cm of b2] (n1tr);
\vertex [below=0.5cm of b3] (n2tl);
\vertex [below=0.5cm of b4] (n2tr);
\vertex [below=0.5cm of b6] (n3tl);
\vertex [below=0.5cm of b7] (n3tr);
\vertex [below=0.5cm of b8] (n4tl);
\vertex [below=0.5cm of b9] (n4tr);
\vertex [below=0.5cm of n1tl] (n1bl);
\vertex [below=0.5cm of n1tr] (n1br);
\vertex [below=0.5cm of n2tl] (n2bl);
\vertex [below=0.5cm of n2tr] (n2br);
\vertex [below=0.5cm of n3tl] (n3bl);
\vertex [below=0.5cm of n3tr] (n3br);
\vertex [below=0.5cm of n4tl] (n4bl);
\vertex [below=0.5cm of n4tr] (n4br);
\vertex [below left=0.35cm of n1bl] (n1l);
\vertex [below right=0.35cm of n1br] (n1r);
\vertex [below left=0.35cm of n2bl] (n2l);
\vertex [below right=0.35cm of n2br] (n2r);
\vertex [below left=0.35cm of n3bl] (n3l);
\vertex [below right=0.35cm of n3br] (n3r);
\vertex [below left=0.35cm of n4bl] (n4l);
\vertex [below right=0.35cm of n4br] (n4r);
\vertex [above right=0.35cm of n1br] (n12lx);
\vertex [right=0.12cm of n12lx] (n12l);
\vertex [above left=0.35cm of n2bl] (n12rx);
\vertex [left=0.12cm of n12rx] (n12r);
\vertex [above right=0.35cm of n3br] (n34lx);
\vertex [right=0.12cm of n34lx] (n34l);
\vertex [above left=0.35cm of n4bl] (n34rx);
\vertex [left=0.12cm of n34rx] (n34r);

\diagram* {
  (l) -- [quarter right] (b1) -- (b2) -- [edge label'=\(z\)] (b3) -- (b4) -- [edge label'=\(\underline{0}\)] (b5) -- [fermion] (b6) -- (b7) -- (b8) -- (b9) -- [quarter right] (r),
  (l) -- [quarter left] (t1) -- (t2) -- [edge label=\(1-z\)] (t3) -- (t4) -- [anti fermion, edge label=\(\underline{x}\)] (t5) -- [edge label=\(\underline{x}'\)] (t6) -- (t7) -- (t8) -- (t9) -- [quarter left] (r),
  (p1) -- [photon, momentum'=\(q\)] (l),
  (r) -- [photon, momentum'=\(q\)] (p2),
  (m1) -- [photon] (n1tl),
  (m2) -- [photon] (n1tr),
  (m3) -- [photon] (n2tl),
  (m4) -- [photon] (n2tr),
  (m6) -- [photon] (n3tl),
  (m7) -- [photon] (n3tr),
  (m8) -- [photon] (n4tl),
  (m9) -- [photon] (n4tr),
  (n1tl) -- (n1tr) -- [half left] (n1br) -- (n1bl) -- [half left] (n1tl),
  (n2tl) -- (n2tr) -- [half left] (n2br) -- (n2bl) -- [half left] (n2tl),
  (n3tl) -- (n3tr) -- [half left] (n3br) -- (n3bl) -- [half left] (n3tl),
  (n4tl) -- (n4tr) -- [half left] (n4br) -- (n4bl) -- [half left] (n4tl),
  (n1bl) -- [double, double distance=1.0pt] (n1l),
  (n1br) -- [double, double distance=1.0pt] (n1r),
  (n2bl) -- [double, double distance=1.0pt] (n2l),
  (n2br) -- [double, double distance=1.0pt] (n2r),
  (n3bl) -- [double, double distance=1.0pt] (n3l),
  (n3br) -- [double, double distance=1.0pt] (n3r),
  (n4bl) -- [double, double distance=1.0pt] (n4l),
  (n4br) -- [double, double distance=1.0pt] (n4r),
  (n12l) -- [ghost] (n12r),
  (n34l) -- [ghost] (n34r),
};

\draw[fill=black!10] (n1tl) -- (n1tr) arc (90:-90:0.25cm) -- (n1bl) arc (-90:-270:0.25cm);
\draw[fill=black!10] (n2tl) -- (n2tr) arc (90:-90:0.25cm) -- (n2bl) arc (-90:-270:0.25cm);
\draw[fill=black!10] (n3tl) -- (n3tr) arc (90:-90:0.25cm) -- (n3bl) arc (-90:-270:0.25cm);
\draw[fill=black!10] (n4tl) -- (n4tr) arc (90:-90:0.25cm) -- (n4bl) arc (-90:-270:0.25cm);

\diagram* {
  (tt) -- [very thick] (bb),
};
\end{feynman}
\end{tikzpicture}
\caption{Illustration of a coherent (elastic) $\gamma^*$-nucleus scattering in $A_+=0$ light cone gauge. \label{fig:1}}
\end{figure}

In this section coherent virtual photon-nucleus scattering will be reviewed where the scattering is done in a frame where the nucleus is at rest, $P_\mu = \qty(\frac{M}{\sqrt{2}},\frac{M}{\sqrt{2}},\underline{0}) = \qty(P_+,P_-,\underline{P})$, and where the virtual photon has momentum $q_\mu = \qty(q_+,-\frac{Q^2}{2q_+},\underline{0})$. Also in this section we use $A_+=0$ gauge, natural for the picture of elastic dipole-nucleus scattering which controls the process illustrated in Fig.~\ref{fig:1}. In our McLerran-Venugopalan (MV) model of the nucleus the ($q\bar{q}$) dipole scatters elastically on nucleons, having a number density $\rho$ in the nucleus, in a two gluon exchange approximation, where the antiquark is at transverse coordinate $\underline{x}$ and the longitudinal momentum $(1-z)q_+$ while the quark is at transverse coordinate $\underline{0}$ and has $z q_+$ longitudinal momentum. The scattering of the ($q\bar{q}$) dipole with a single nucleon of the target is given by the
\begin{equation} \label{eq:1}
    T\qty(x_\perp) = \frac{\pi^2\alpha C_F}{N_C^2-1} xG\qty(x,x_\perp^2) x_\perp^2
\end{equation}
with $xG$ the nuclon gluon distribution at a scale $\frac{1}{x_\perp^2}$, where \eqref{eq:1} corresponds to a quark saturation momentum $Q_S$ given as
\begin{equation} \label{eq:2_}
    Q_S^2 = \frac{4\pi^2\alpha C_F}{N_C^2-1} \rho L\, xG
\end{equation}
with $xG$ in \eqref{eq:2_} evaluated at $x_\perp^2 = \frac{4}{Q_S^2}$. $L$ is the length of the nucleus at the impact parameter in question while $\rho$ is the nuclear density.

Then the coherent scattering (elastic dipole-nucleus scattering) illustrated in Fig.~\ref{fig:1} \cite{3} and with $\bar{Q}^2 = Q^2z(1-z)$ is
\begin{equation} \label{eq:2}
    \frac{\dd{\sigma_T^{\gamma^*}}}{\dd[2]{b}\dd[2]{p}\dd{z}} = 2\alpha_{em} N_C e_f^2 \qty[z^2 + (1-z)^2] A
\end{equation}
\begin{equation} \label{eq:3}
    A = \bar{Q}^2 \int \frac{\dd[2]{x}\dd[2]{x'}}{\qty(2\pi)^4} e^{i\underline{p}{\cdot}\qty(\underline{x}-\underline{x'})} K_1\qty(\bar{Q}x_\perp) \frac{\underline{x}{\cdot}\underline{x'}}{x_\perp x'_\perp} K_1\qty(\bar{Q}x'_\perp) T\qty(x_\perp) T\qty(x'_\perp)
\end{equation}
where the elastic dipole-nucleus scattering amplitude is
\begin{equation} \label{eq:4}
    T(b,x_\perp) = 1-e^{-Q_S^2 x_\perp^2/4} = 1 - S(b,x_\perp) \,.
\end{equation}
The exponentiation in \eqref{eq:4} is manifest in either a covariant gauge or in an $A_+=0$ gauge where the scattering of the dipole off the nucleons is sequential.

\subsection{Inelastic scattering}
\label{sec:2.2}

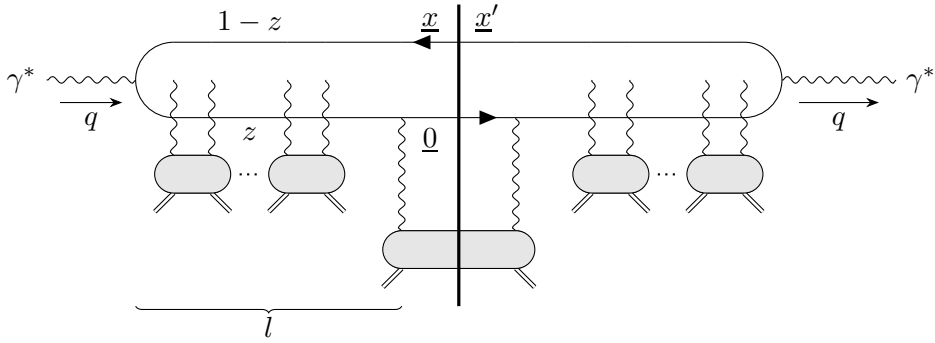
\begin{figure}[htbp]
\centering
\begin{tikzpicture}
\begin{feynman} 
\vertex (p1) {\(\gamma^*\)};
\vertex [right=of p1] (l);
\vertex [below right=0.707cm of l] (b1);
\vertex [right=0.5cm of b1] (b2);
\vertex [right=1.0cm of b2] (b3);
\vertex [right=0.5cm of b3] (b4);
\vertex [right=1.0cm of b4] (b5);
\vertex [right=0.75cm of b5] (b6);
\vertex [right=0.75cm of b6] (b7);
\vertex [right=1.0cm of b7] (b8);
\vertex [right=0.5cm of b8] (b9);
\vertex [right=1.0cm of b9] (b10);
\vertex [right=0.5cm of b10] (b11);
\vertex [above right=0.707cm of b11] (r);
\vertex [right=of r] (p2) {\(\gamma^*\)};
\vertex [above right=0.707cm of l] (t1);
\vertex [right=0.5cm of t1] (t2);
\vertex [right=1.0cm of t2] (t3);
\vertex [right=0.5cm of t3] (t4);
\vertex [right=1.0cm of t4] (t5);
\vertex [right=0.75cm of t5] (t6);
\vertex [right=0.75cm of t6] (t7);
\vertex [right=1.0cm of t7] (t8);
\vertex [right=0.5cm of t8] (t9);
\vertex [right=1.0cm of t9] (t10);
\vertex [right=0.5cm of t10] (t11);
\vertex [right=0.5cm of l] (m1);
\vertex [right=0.5cm of m1] (m2);
\vertex [right=1.0cm of m2] (m3);
\vertex [right=0.5cm of m3] (m4);
\vertex [right=1.0cm of m4] (m5);
\vertex [right=0.75cm of m5] (m6);
\vertex [right=0.75cm of m6] (m7);
\vertex [right=1.0cm of m7] (m8);
\vertex [right=0.5cm of m8] (m9);
\vertex [right=1.0cm of m9] (m10);
\vertex [right=0.5cm of m10] (m11);
\vertex [above=0.5cm of t6] (tt);
\vertex [below=2.5cm of b6] (bb);
\vertex [below=0.5cm of b1] (n1tl);
\vertex [below=0.5cm of b2] (n1tr);
\vertex [below=0.5cm of b3] (n2tl);
\vertex [below=0.5cm of b4] (n2tr);
\vertex [below=of b5] (nctl);
\vertex [below=of b7] (nctr);
\vertex [below=0.5cm of b8] (n3tl);
\vertex [below=0.5cm of b9] (n3tr);
\vertex [below=0.5cm of b10] (n4tl);
\vertex [below=0.5cm of b11] (n4tr);
\vertex [below=0.5cm of n1tl] (n1bl);
\vertex [below=0.5cm of n1tr] (n1br);
\vertex [below=0.5cm of n2tl] (n2bl);
\vertex [below=0.5cm of n2tr] (n2br);
\vertex [below=0.5cm of nctl] (ncbl);
\vertex [below=0.5cm of nctr] (ncbr);
\vertex [below=0.5cm of n3tl] (n3bl);
\vertex [below=0.5cm of n3tr] (n3br);
\vertex [below=0.5cm of n4tl] (n4bl);
\vertex [below=0.5cm of n4tr] (n4br);
\vertex [below left=0.35cm of n1bl] (n1l);
\vertex [below right=0.35cm of n1br] (n1r);
\vertex [below left=0.35cm of n2bl] (n2l);
\vertex [below right=0.35cm of n2br] (n2r);
\vertex [below left=0.35cm of ncbl] (ncl);
\vertex [below right=0.35cm of ncbr] (ncr);
\vertex [below left=0.35cm of n3bl] (n3l);
\vertex [below right=0.35cm of n3br] (n3r);
\vertex [below left=0.35cm of n4bl] (n4l);
\vertex [below right=0.35cm of n4br] (n4r);
\vertex [right=0.12cm of n12lx] (n12l);
\vertex [above left=0.35cm of n2bl] (n12rx);
\vertex [left=0.12cm of n12rx] (n12r);
\vertex [above right=0.35cm of n3br] (n34lx);
\vertex [right=0.12cm of n34lx] (n34l);
\vertex [above left=0.35cm of n4bl] (n34rx);
\vertex [left=0.12cm of n34rx] (n34r);
\vertex [below=3.0cm of l] (brl);
\vertex [below=2.5cm of b5] (brr);

\diagram* {
  (l) -- [quarter right] (b1) -- (b2) -- [edge label'=\(z\)] (b3) -- (b4) -- (b5) -- [edge label'=\(\underline{0}\)] (b6) -- [fermion] (b7) -- (b8) -- (b9) -- (b10) -- (b11) -- [quarter right] (r),
  (l) -- [quarter left] (t1) -- (t2) -- [edge label=\(1-z\)] (t3) -- (t4) -- (t5) -- [anti fermion, edge label=\(\underline{x}\)] (t6) -- [edge label=\(\underline{x}'\)] (t7) -- (t8) -- (t9) -- (t10) -- (t11) -- [quarter left] (r),
  (p1) -- [photon, momentum'=\(q\)] (l),
  (r) -- [photon, momentum'=\(q\)] (p2),
  (m1) -- [photon] (n1tl),
  (m2) -- [photon] (n1tr),
  (m3) -- [photon] (n2tl),
  (m4) -- [photon] (n2tr),
  (b5) -- [photon] (nctl),
  (b7) -- [photon] (nctr),
  (m8) -- [photon] (n3tl),
  (m9) -- [photon] (n3tr),
  (m10) -- [photon] (n4tl),
  (m11) -- [photon] (n4tr),
  (n1tl) -- (n1tr) -- [half left] (n1br) -- (n1bl) -- [half left] (n1tl),
  (n2tl) -- (n2tr) -- [half left] (n2br) -- (n2bl) -- [half left] (n2tl),
  (nctl) -- (nctr) -- [half left] (ncbr) -- (ncbl) -- [half left] (nctl),
  (n3tl) -- (n3tr) -- [half left] (n3br) -- (n3bl) -- [half left] (n3tl),
  (n4tl) -- (n4tr) -- [half left] (n4br) -- (n4bl) -- [half left] (n4tl),
  (n1bl) -- [double, double distance=1.0pt] (n1l),
  (n1br) -- [double, double distance=1.0pt] (n1r),
  (n2bl) -- [double, double distance=1.0pt] (n2l),
  (n2br) -- [double, double distance=1.0pt] (n2r),
  (ncbl) -- [double, double distance=1.0pt] (ncl),
  (ncbr) -- [double, double distance=1.0pt] (ncr),
  (n3bl) -- [double, double distance=1.0pt] (n3l),
  (n3br) -- [double, double distance=1.0pt] (n3r),
  (n4bl) -- [double, double distance=1.0pt] (n4l),
  (n4br) -- [double, double distance=1.0pt] (n4r),
  (n12l) -- [ghost] (n12r),
  (n34l) -- [ghost] (n34r),
};

\draw[fill=black!10] (n1tl) -- (n1tr) arc (90:-90:0.25cm) -- (n1bl) arc (-90:-270:0.25cm);
\draw[fill=black!10] (n2tl) -- (n2tr) arc (90:-90:0.25cm) -- (n2bl) arc (-90:-270:0.25cm);
\draw[fill=black!10] (nctl) -- (nctr) arc (90:-90:0.25cm) -- (ncbl) arc (-90:-270:0.25cm);
\draw[fill=black!10] (n3tl) -- (n3tr) arc (90:-90:0.25cm) -- (n3bl) arc (-90:-270:0.25cm);
\draw[fill=black!10] (n4tl) -- (n4tr) arc (90:-90:0.25cm) -- (n4bl) arc (-90:-270:0.25cm);
\draw [decoration={brace}, decorate] (brr.south) -- (brl.south)
node [pos=0.5, below] {\(l\)};

\diagram* {
  (tt) -- [very thick] (bb),
};
\end{feynman}
\end{tikzpicture}
\caption{Illustration of an inelastic $\gamma^*$-nucleus scattering in $A_+=0$ light cone gauge. \label{fig:2}}
\end{figure}

An inelastic scattering event is shown in Fig.~\ref{fig:2}. Now the dipole has only elastic scatterings, in both the amplitude and complex conjugate amplitude, up to a distance $l$ in the nucleus at which point it has an inelastic reaction with a nucleon of the nucleus. After the inelastic scattering with a nucleon the total inelastic cross section does not depend on interactions which happen at distances between $l$ and $L$ so they are ignored in this discussion. Calling $\hat{q} = \frac{Q_S^2}{L}$,
\begin{equation} \label{eq:5}
    \sigma_{in}(b) = 2 \alpha_{em} N_C e_f^2 \bar{Q}^2 \int \frac{\dd[2]{x}}{4\pi^2} \qty[z^2 + (1-z)^2] \dd{z} K_1^2\qty(\bar{Q}x_\perp) \int_0^L \dd{l} e^{-2l\hat{q}x_\perp^2/4} \frac{2\hat{q}x_\perp^2}{4}
\end{equation}
where the exponential factor in \eqref{eq:5} represents the probability of not having an inelastic reaction, in the amplitude and complex conjugate amplitude, up to a distance $l$ in the nucleus while $2\hat{q}x_\perp^2/4 \dd{l}$ represents the probability of having an inelastic reaction in the interval $\dd{l}$. One finds \cite{3}
\begin{equation} \label{eq:6}
    \sigma_{in}(b) = 2 \alpha_{em} N_C e_f^2 \bar{Q}^2 \int \frac{\dd[2]{x}}{4\pi^2} \qty[z^2 + (1-z)^2] K_1^2\qty(\bar{Q}x_\perp) \qty[1-S^2(b,x_\perp] \dd{z} \,.
\end{equation}

Equations \eqref{eq:3} and \eqref{eq:6} are well-known. In the region $Q_S^2x_\perp^2/4 \gg 1$, $\sigma_{in} = \sigma_{el}$ and one says that the scattering is in the (quark) saturation regime which, from the projectile point of view, simply means that the unitary limits for elastic and inelastic scattering have been reached. It is widely believed that when scattering occurs in the unitarity regime it should correspond, from the target point of view, to a maximum value of quark (and antiquark) occupations of the nuclear wave function and that this should be manifest in an $A_-=0$ gauge calculation. We now turn to exactly this issue.

\section{Projectile evolution in $A_-=0$ gauge}
\label{sec:3}

\subsection{Coherent single scattering; first version}
\label{sec:3.1}

\begin{figure}[htbp]
\centering
\begin{tikzpicture}
\begin{feynman} 
\vertex (p1);
\vertex [right=of p1] (l);
\vertex [below right=1cm of l] (b1);
\vertex [right=1.0cm of b1] (b2);
\vertex [right=0.5cm of b2] (b3);
\vertex [right=1.5cm of b3] (b4);
\vertex [right=0.5cm of b4] (b5);
\vertex [right=1.0cm of b5] (b6);
\vertex [above right=1cm of l] (t1);
\vertex [right=1.0cm of t1] (t2);
\vertex [right=0.5cm of t2] (t3);
\vertex [right=1.5cm of t3] (t4);
\vertex [right=0.5cm of t4] (t5);
\vertex [right=1.0cm of t5] (t6);
\vertex [below=1.5cm of b3] (ntl);
\vertex [below=1.5cm of b4] (ntr);
\vertex [below=0.75cm of ntl] (nbl);
\vertex [below=0.75cm of ntr] (nbr);
\vertex [below left=0.7cm of nbl] (nl);
\vertex [below right=0.7cm of nbr] (nr);
\vertex [above=0.02cm of b3] {\(\alpha\)};
\vertex [above left=0.05cm of ntl] {\(\beta\)};

\diagram* {
  (l) -- [quarter right] (b1) -- [edge label=\(z\), momentum'=\(p\)] (b2) -- (b3) -- [fermion, momentum=\(p+l\)] (b4) -- (b5) -- [momentum=\(p+\delta\)] (b6),
  (l) -- [quarter left] (t1) -- [edge label=\(1-z\)] (t2) -- (t3) -- [anti fermion] (t4) -- (t5) -- [momentum=\(q-p\)] (t6),
  (p1) -- [photon, momentum'=\(q\)] (l),
  (ntl) -- [photon, momentum'=\(l\)] (b3),
  (b4) -- [photon, momentum=\(l-\delta\)] (ntr),
  (ntl) -- (ntr) -- [half left] (nbr) -- (nbl) -- [half left] (ntl),
  (nl) -- [double, double distance=1.0pt, momentum=\(P\)] (nbl),
  (nbr) -- [double, double distance=1.0pt, momentum=\(P-\delta\)] (nr),
};

\draw[fill=black!10] (ntl) -- (ntr) arc (90:-90:0.375cm) -- (nbl) arc (-90:-270:0.375cm);
\end{feynman}
\end{tikzpicture}
\caption{A dipole nucleon elastic scattering in $A_-=0$ light cone gauge. \label{fig:3}}
\end{figure}
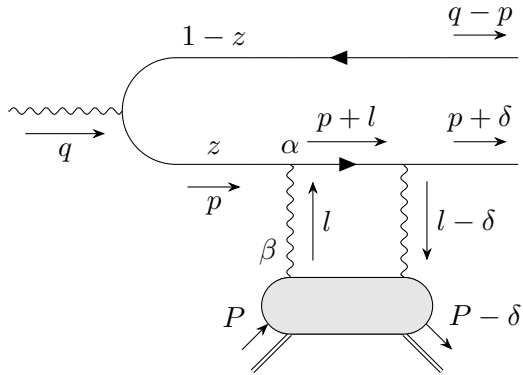

The picture in Fig.~\ref{fig:1} no longer applies when one calculates $\sigma_{el}$ in $A_-=0$ gauge because the sequential nature of the multiple scatterings is not manifest in this gauge. By working through a lowest order example one can see how to proceed in this gauge. It is convenient at the beginning to consider just the amplitude and to do the evaluation completely in momentum space. Of the various momenta we choose $P$ and $q$ as in Sec.~\ref{sec:2} and label $p$ and $\delta$ as
\begin{equation} \label{eq:7}
    p_\mu \simeq \qty(p_+,-\frac{Q^2}{2q_+},\underline{p})
\end{equation}
\begin{equation} \label{eq:8}
    \delta_\mu \simeq \qty(0,\frac{Q^2}{2q_+} + \frac{p^2}{2p_+},\underline{0}) \,,
\end{equation}
and we suppose $\frac{\underline{p}^2}{Q^2},\frac{\underline{l}^2}{\underline{p}^2} \ll 1$ valid in the logarithmic regions of integration for $p_\perp$ and $l_\perp$. Lines $q-p$ and $p+\delta$ are on-shell. The process and kinematics are shown in Fig.~\ref{fig:3}.

We focus first on the $l_-$-integration with the $A_-=0$ gluon propagator shown in Fig.~\ref{fig:4} given as
\begin{equation} \label{eq:9}
    G_{\beta\alpha}(l) = \frac{-i}{l^2+i\epsilon} \qty[g_{\beta\alpha} - \frac{\eta_\beta l_\alpha}{l_--i\epsilon} - \frac{\eta_\alpha l_\beta}{l_-+i\epsilon}] \,,
\end{equation}
when $\eta{\cdot}\nu = \nu_-$ for any vector $\nu_\mu$. Choices of $i\epsilon$ different than that of \eqref{eq:9} can be taken but are more delicate and do not lead so quickly to the desired form of the final result.

\begin{figure}[htbp]
\centering
\feynmandiagram [horizontal=a to b] {
  a [particle=\(\beta\)] -- [photon, momentum'=\(l\)] b [particle=\(\alpha\)],
};
\caption{The gluon propagator, $G_{\beta\alpha}(l)$. \label{fig:4}}
\end{figure}

\begin{figure}[htbp]
\centering
\begin{subfigure}[b]{0.49\textwidth}
\centering
\begin{tikzpicture}
\begin{feynman} 
\vertex (p1);
\vertex [right=of p1] (l);
\vertex [below right=1cm of l] (b1);
\vertex [right=1.0cm of b1] (b2);
\vertex [right=0.5cm of b2] (b3);
\vertex [right=1.5cm of b3] (b4);
\vertex [right=0.5cm of b4] (b5);
\vertex [right=1.0cm of b5] (b6);
\vertex [above right=1cm of l] (t1);
\vertex [right=1.0cm of t1] (t2);
\vertex [right=0.5cm of t2] (t3);
\vertex [right=1.5cm of t3] (t4);
\vertex [right=0.5cm of t4] (t5);
\vertex [right=1.0cm of t5] (t6);
\vertex [below=1.5cm of b3] (ntl);
\vertex [below=1.5cm of b4] (ntr);
\vertex [below=0.75cm of ntl] (nbl);
\vertex [below=0.75cm of ntr] (nbr);
\vertex [below left=0.7cm of nbl] (nl);
\vertex [below right=0.7cm of nbr] (nr);

\diagram* {
  (l) -- [quarter right] (b1) -- [momentum=\(p+l\)] (b2) -- (b3) -- [fermion] (b4) -- (b5) -- [momentum=\(p+\delta\)] (b6),
  (l) -- [quarter left] (t1) -- (t2) -- (t3) -- [anti fermion] (t4) -- (t5) -- [momentum=\(q-p\)] (t6),
  (p1) -- [photon, momentum'=\(q\)] (l),
  (ntl) -- [scalar, quarter left, momentum=\(l\)] (l),
  (b4) -- [photon, momentum=\(l-\delta\)] (ntr),
  (ntl) -- (ntr) -- [half left] (nbr) -- (nbl) -- [half left] (ntl),
  (nl) -- [double, double distance=1.0pt, momentum=\(P\)] (nbl),
  (nbr) -- [double, double distance=1.0pt, momentum=\(P-\delta\)] (nr),
};

\draw[fill=black!10] (ntl) -- (ntr) arc (90:-90:0.375cm) -- (nbl) arc (-90:-270:0.375cm);
\end{feynman}
\end{tikzpicture}
\caption{\label{fig:5A}}
\end{subfigure}
\hfill
\begin{subfigure}[b]{0.49\textwidth}
\centering
\begin{tikzpicture}
\begin{feynman} 
\vertex (p1);
\vertex [right=of p1] (l);
\vertex [below right=1cm of l] (b1);
\vertex [right=1.0cm of b1] (b2);
\vertex [right=0.5cm of b2] (b3);
\vertex [right=1.5cm of b3] (b4);
\vertex [right=0.5cm of b4] (b5);
\vertex [right=1.0cm of b5] (b6);
\vertex [above right=1cm of l] (t1);
\vertex [right=1.0cm of t1] (t2);
\vertex [right=0.5cm of t2] (t3);
\vertex [right=1.5cm of t3] (t4);
\vertex [right=0.5cm of t4] (t5);
\vertex [right=1.0cm of t5] (t6);
\vertex [below=1.5cm of b3] (ntl);
\vertex [below=1.5cm of b4] (ntr);
\vertex [below=0.75cm of ntl] (nbl);
\vertex [below=0.75cm of ntr] (nbr);
\vertex [below left=0.7cm of nbl] (nl);
\vertex [below right=0.7cm of nbr] (nr);

\diagram* {
  (l) -- [quarter right] (b1) -- [momentum=\(p\)] (b2) -- (b3) -- [fermion] (b4) -- (b5) -- [momentum=\(p+\delta\)] (b6),
  (l) -- [quarter left] (t1) -- (t2) -- (t3) -- [anti fermion] (t4) -- (t5) -- [momentum=\(q-p\)] (t6),
  (p1) -- [photon, momentum'=\(q\)] (l),
  (ntl) -- [scalar, quarter left, momentum'=\(l\)] (b4),
  (b4) -- [photon, momentum=\(l-\delta\)] (ntr),
  (ntl) -- (ntr) -- [half left] (nbr) -- (nbl) -- [half left] (ntl),
  (nl) -- [double, double distance=1.0pt, momentum=\(P\)] (nbl),
  (nbr) -- [double, double distance=1.0pt, momentum=\(P-\delta\)] (nr),
};

\draw[fill=black!10] (ntl) -- (ntr) arc (90:-90:0.375cm) -- (nbl) arc (-90:-270:0.375cm);
\end{feynman}
\end{tikzpicture}
\caption{\label{fig:5B}}
\end{subfigure}
\caption{Using the relation \eqref{eq:11} in Fig.~\ref{fig:3} leads to the result A$-$B with A and B illustrated with the relevant kinematics. \label{fig:5}}
\end{figure}
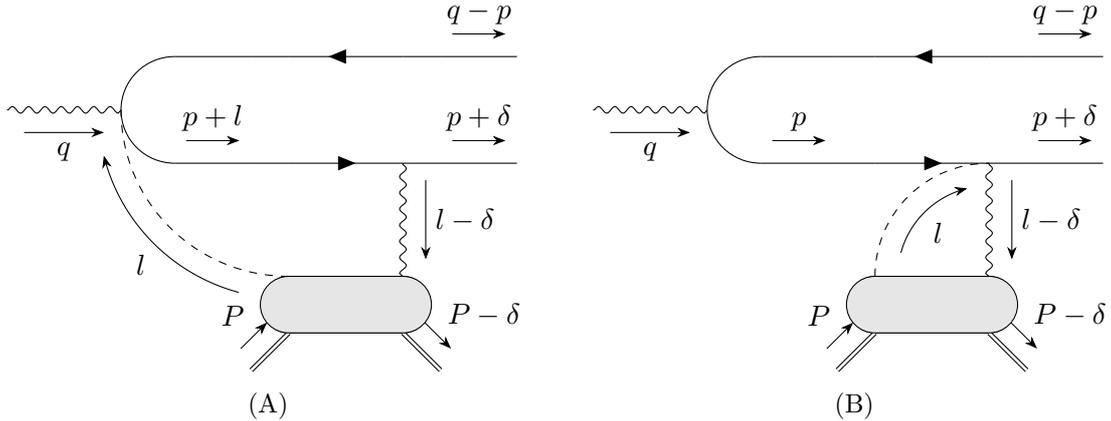

Denominators involving $l_-$ come from the $l$,$l-\delta$ and $(p+l)$ propagators with $l_-$ and $(l-\delta)_-$ from the $l$ and $(l-\delta)$ propagators. The $l_-$-factors are then
\begin{equation} \label{eq:10}
    \frac{1}{\qty(l_--i\epsilon) \qty[(l-\delta)_-+i\epsilon]^2}
\end{equation}
where $\delta_- = \frac{Q^2}{2q_+}+\frac{p^2}{2p_+}$ is fixed by $(p+\delta)^2=0$. Thus one may do the $l_-$-integration by distorting the $l_-$-integration over the pole at $l_-=i\epsilon$. (Only the $\frac{1}{l_--i\epsilon}$ and not the $\frac{1}{l_-+i\epsilon}$ from \eqref{eq:9} contributes because the $\eta_\alpha$ from \eqref{eq:9} gives a very small result when used in the graph shown in Fig.~\ref{fig:3}. Also the $l$-line only couples to the $p$-line and not the $(q-p)$-line in that figure if we take the logarithmic $p_\perp$ integration to correspond to $z$ small and $(1-z)$ near 1. There of course is an equivalent region where $(1-z)$ is small and $z$ large where $l$ hooks onto the $(q-p)$-line.) Once the contour integration is done to get $l_-=0$, and noticing that $l_+\simeq 0$, one sees that for the dominant $\eta_\beta l_\alpha^\perp$ term of \eqref{eq:9}, when used in Fig.~\ref{fig:3}, $l_\alpha^\perp \simeq l_\alpha$ and the $l_\alpha$ can be used to generate Ward identities for the graph in Fig.~\ref{fig:3}. That is, referring to Fig.~\ref{fig:3}, one writes
\begin{equation} \label{eq:11}
    il_\alpha \frac{i}{\gamma{\cdot}(p+l)} \gamma_\alpha \frac{i}{\gamma{\cdot} p} = \frac{i}{\gamma{\cdot}(p+l)} - \frac{i}{\gamma{\cdot} p}
\end{equation}
so that Fig.~\ref{fig:3} gets replaced by Fig.~\ref{fig:5} where the dashed line is that figure carries color and transverse momentum but has a scalar Lorentz structure. Graph B in Fig.~\ref{fig:5} cannot have an $l_\perp^2$ in the numerator, necessary to get the gluon distribution of the nucleon, $P$, so we drop this graph and only keep graph A.

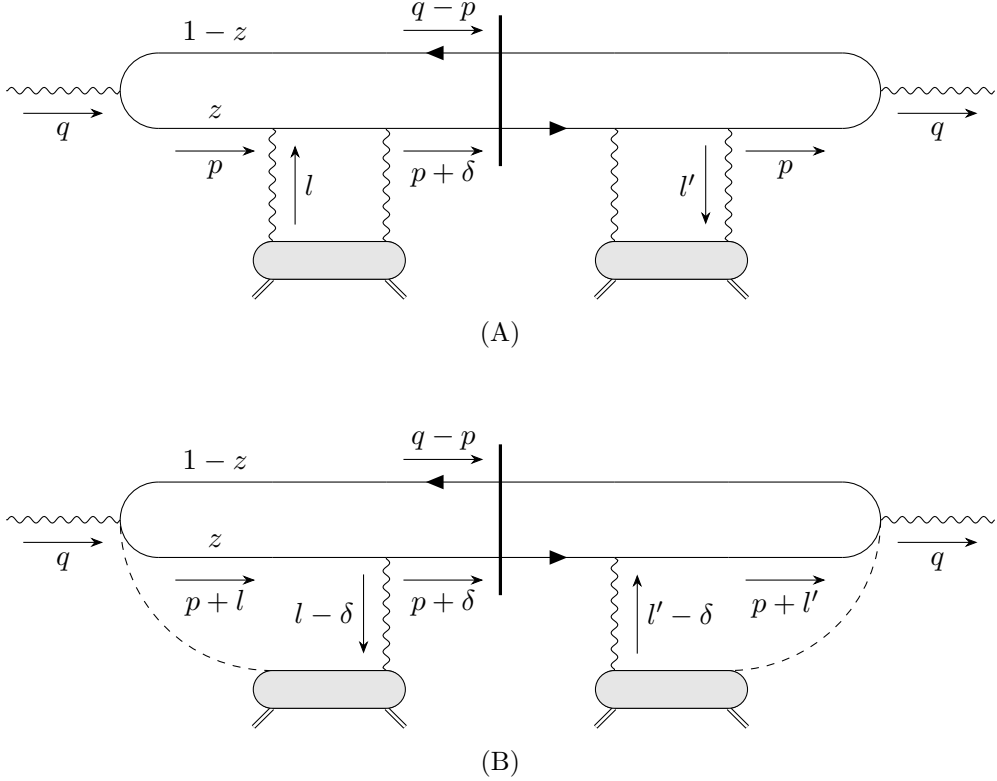
\begin{figure}[htbp]
\centering
\begin{subfigure}[b]{0.99\textwidth}
\centering
\begin{tikzpicture}
\begin{feynman} 
\vertex (p1);
\vertex [right=of p1] (l);
\vertex [below right=0.707cm of l] (b1);
\vertex [right=of b1] (b2);
\vertex [right=of b2] (b3);
\vertex [right=of b3] (b4);
\vertex [right=of b4] (b5);
\vertex [right=of b5] (b6);
\vertex [right=of b6] (b7);
\vertex [above right=0.707cm of b7] (r);
\vertex [right=of r] (p2);
\vertex [above right=0.707cm of l] (t1);
\vertex [right=of t1] (t2);
\vertex [right=of t2] (t3);
\vertex [right=of t3] (t4);
\vertex [right=of t4] (t5);
\vertex [right=of t5] (t6);
\vertex [right=of t6] (t7);
\vertex [above=0.5cm of t4] (tt);
\vertex [below=0.5cm of b4] (bb);
\vertex [below= of b2] (n1tl);
\vertex [below= of b3] (n1tr);
\vertex [below= of b5] (n2tl);
\vertex [below= of b6] (n2tr);
\vertex [below=0.5cm of n1tl] (n1bl);
\vertex [below=0.5cm of n1tr] (n1br);
\vertex [below=0.5cm of n2tl] (n2bl);
\vertex [below=0.5cm of n2tr] (n2br);
\vertex [below left=0.35cm of n1bl] (n1l);
\vertex [below right=0.35cm of n1br] (n1r);
\vertex [below left=0.35cm of n2bl] (n2l);
\vertex [below right=0.35cm of n2br] (n2r);

\diagram* {
  (l) -- [quarter right] (b1) -- [edge label=\(z\), momentum'=\(p\)] (b2) -- (b3) -- [momentum'=\(p+\delta\)] (b4) -- [fermion] (b5) -- (b6) -- [momentum'=\(p\)] (b7) -- [quarter right] (r),
  (l) -- [quarter left] (t1) -- [edge label=\(1-z\)] (t2) -- (t3) -- [anti fermion, momentum=\(q-p\)] (t4) -- (t5) -- (t6) -- (t7) -- [quarter left] (r),
  (p1) -- [photon, momentum'=\(q\)] (l),
  (r) -- [photon, momentum'=\(q\)] (p2),
  (n1tl) -- [photon, momentum'=\(l\)] (b2),
  (b3) -- [photon] (n1tr),
  (n2tl) -- [photon] (b5),
  (b6) -- [photon, momentum'=\(l'\)] (n2tr),
  (n1tl) -- (n1tr) -- [half left] (n1br) -- (n1bl) -- [half left] (n1tl),
  (n2tl) -- (n2tr) -- [half left] (n2br) -- (n2bl) -- [half left] (n2tl),
  (n1l) -- [double, double distance=1.0pt] (n1bl),
  (n1br) -- [double, double distance=1.0pt] (n1r),
  (n2l) -- [double, double distance=1.0pt] (n2bl),
  (n2br) -- [double, double distance=1.0pt] (n2r),
};

\draw[fill=black!10] (n1tl) -- (n1tr) arc (90:-90:0.25cm) -- (n1bl) arc (-90:-270:0.25cm);
\draw[fill=black!10] (n2tl) -- (n2tr) arc (90:-90:0.25cm) -- (n2bl) arc (-90:-270:0.25cm);

\diagram* {
  (tt) -- [very thick] (bb),
};
\end{feynman}
\end{tikzpicture}
\caption{\label{fig:6A}}
\end{subfigure}
\par\bigskip
\par\bigskip
\begin{subfigure}[b]{0.99\textwidth}
\centering
\begin{tikzpicture}
\begin{feynman} 
\vertex (p1);
\vertex [right=of p1] (l);
\vertex [below right=0.707cm of l] (b1);
\vertex [right=of b1] (b2);
\vertex [right=of b2] (b3);
\vertex [right=of b3] (b4);
\vertex [right=of b4] (b5);
\vertex [right=of b5] (b6);
\vertex [right=of b6] (b7);
\vertex [above right=0.707cm of b7] (r);
\vertex [right=of r] (p2);
\vertex [above right=0.707cm of l] (t1);
\vertex [right=of t1] (t2);
\vertex [right=of t2] (t3);
\vertex [right=of t3] (t4);
\vertex [right=of t4] (t5);
\vertex [right=of t5] (t6);
\vertex [right=of t6] (t7);
\vertex [above=0.5cm of t4] (tt);
\vertex [below=0.5cm of b4] (bb);
\vertex [below= of b2] (n1tl);
\vertex [below= of b3] (n1tr);
\vertex [below= of b5] (n2tl);
\vertex [below= of b6] (n2tr);
\vertex [below=0.5cm of n1tl] (n1bl);
\vertex [below=0.5cm of n1tr] (n1br);
\vertex [below=0.5cm of n2tl] (n2bl);
\vertex [below=0.5cm of n2tr] (n2br);
\vertex [below left=0.35cm of n1bl] (n1l);
\vertex [below right=0.35cm of n1br] (n1r);
\vertex [below left=0.35cm of n2bl] (n2l);
\vertex [below right=0.35cm of n2br] (n2r);

\diagram* {
  (l) -- [quarter right] (b1) -- [edge label=\(z\), momentum'=\(p+l\)] (b2) -- (b3) -- [momentum'=\(p+\delta\)] (b4) -- [fermion] (b5) -- (b6) -- [momentum'=\(p+l'\)] (b7) -- [quarter right] (r),
  (l) -- [quarter left] (t1) -- [edge label=\(1-z\)] (t2) -- (t3) -- [anti fermion, momentum=\(q-p\)] (t4) -- (t5) -- (t6) -- (t7) -- [quarter left] (r),
  (p1) -- [photon, momentum'=\(q\)] (l),
  (r) -- [photon, momentum'=\(q\)] (p2),
  (n1tl) -- [scalar, quarter left] (l),
  (b3) -- [photon, momentum'=\(l-\delta\)] (n1tr),
  (n2tl) -- [photon, momentum'=\(l'-\delta\)] (b5),
  (r) -- [scalar, quarter left] (n2tr),
  (n1tl) -- (n1tr) -- [half left] (n1br) -- (n1bl) -- [half left] (n1tl),
  (n2tl) -- (n2tr) -- [half left] (n2br) -- (n2bl) -- [half left] (n2tl),
  (n1l) -- [double, double distance=1.0pt] (n1bl),
  (n1br) -- [double, double distance=1.0pt] (n1r),
  (n2l) -- [double, double distance=1.0pt] (n2bl),
  (n2br) -- [double, double distance=1.0pt] (n2r),
};

\draw[fill=black!10] (n1tl) -- (n1tr) arc (90:-90:0.25cm) -- (n1bl) arc (-90:-270:0.25cm);
\draw[fill=black!10] (n2tl) -- (n2tr) arc (90:-90:0.25cm) -- (n2bl) arc (-90:-270:0.25cm);

\diagram* {
  (tt) -- [very thick] (bb),
};
\end{feynman}
\end{tikzpicture}
\caption{\label{fig:6B}}
\end{subfigure}
\caption{Graph A shows the term being evaluated which becomes graph B after using the Ward identities. \label{fig:6}}
\end{figure}

Now include the complex conjugate amplitude for single scattering as illustrated in Fig.~\ref{fig:6}. Let's evaluate the Fig.~\ref{fig:6B} version of this graph, which emerge after using the Ward identities. The trace factor then is
\begin{equation} \label{eq:12}
    \tr = \tr \underline{\gamma} {\cdot} \underline{\epsilon}_\gamma^*\ \gamma{\cdot}(p+l')\ \underline{\gamma}{\cdot}\underline{l}'\ \gamma{\cdot}(p+\delta)\ \underline{\gamma}{\cdot}\underline{l}\ \gamma{\cdot}(p+l)\ \underline{\gamma}{\cdot}\underline{\epsilon}_\gamma\ \gamma{\cdot}(q-p) \,.
\end{equation}
Using $\underline{\gamma} {\cdot} \underline{\epsilon}_\gamma^*\ \underline{\gamma} {\cdot} \underline{\epsilon}_\gamma = -1$ and 
$\gamma{\cdot}(q-p) \simeq \gamma_- q_+$ one finds
\begin{equation} \label{eq:13}
    \tr = \tr_{ll'} + \tr_l + \tr_{l'} + \tr_0
\end{equation}
with
\begin{equation} \label{eq:14}
    \tr_{ll'} = 4q_+ \underline{l}^2 \underline{l}^{\prime 2} (p+\delta)_-
\end{equation}
\begin{equation} \label{eq:15}
    \tr_l = 4q_+ \underline{l}^2 \underline{p}{\cdot}\underline{l}' \qty[p_-+(p+\delta)_-]
\end{equation}
\begin{equation} \label{eq:16}
    \tr_{l'} = 4q_+ \underline{l}^{\prime 2} \underline{p}{\cdot}\underline{l} \qty[p_-+(p+\delta)_-]
\end{equation}
\begin{equation} \label{eq:17}
    \tr_0 = 4q_+ \qty[\underline{l}{\cdot}\underline{l}' \qty(2p_-^2p_+ - (p+\delta)_-\underbar{p}^2) + 4p_- \underline{p}{\cdot}\underline{l}\ \underline{p}{\cdot}\underline{l}']
\end{equation}

Next evaluate the denominators $\frac{1}{(p+l)^2}$ and $\frac{1}{(p+l')^2}$ in the graph shown in Fig.~\ref{fig:6B}. One finds
\begin{equation} \label{eq:18}
    \frac{1}{(p+l)^2(p+l')^2} \simeq \frac{1}{\qty(\underline{p}^2 + \bar{Q}^2)^2} \qty(1-\frac{2\underline{p}{\cdot}\underline{l}}{\underline{p}^2+\bar{Q}^2}) \qty(1-\frac{2\underline{p}{\cdot}\underline{l}'}{\underline{p}^2+\bar{Q}^2}) \,.
\end{equation}
In multiplying \eqref{eq:18} times the trace given in \eqref{eq:13}-\eqref{eq:17} one keeps quadratic terms in $\underline{l}$ and $\underline{l}'$. In addition there are factors of $\frac{1}{\delta_-^2}$ from the light cone denominators of the $(l-\delta)$ and $(l'-\delta)$ lines as well as a factor $\frac{1}{2(q-k)_+2p_+} \simeq \frac{1}{4q_+p_+}$ for the external state factors of the $(q-p)$ and $(p+\delta)$-lines. Averaging over the angles of $\underline{l}$ and $\underline{l}'$ one gets
\begin{equation} \label{eq:19}
    A = \frac{\underline{l}^2 \underline{l}^2 \underline{p}^2 \qty[\bar{Q}^2]^2}{\pi^2 \qty[\underline{p}^2 + \bar{Q}^2]^6} \,.
\end{equation}
Before trying to compare \eqref{eq:19} to \eqref{eq:3} one must first write \eqref{eq:19} in a coordinate space expression which is easy to do starting with
\begin{equation} \label{eq:20}
    \frac{\underline{p}}{\qty[\underline{p}^2 + \bar{Q}^2]^3} = \frac{-i}{16\pi\bar{Q}} \int \dd[2]{x} \underline{x} x_\perp K_1\qty(\bar{Q}x_\perp) e^{-i\underline{p}{\cdot}\underline{x}} \,.
\end{equation}

Taking the square of \eqref{eq:20} gives
\begin{equation} \label{eq:21}
    A = \frac{\underline{l}^2 \underline{l}^2 \underline{p}^2 \qty[\bar{Q}^2]^2}{\pi^2 \qty[\underline{p}^2 + \bar{Q}^2]^6} = \frac{1}{16} \bar{Q}^2 \int \frac{\dd[2]{x}\dd[2]{x'}}{(2\pi)^4} e^{-i\underline{p}{\cdot}\qty(\underline{x}-\underline{x}')} K_1\qty(\bar{Q}x_\perp) \frac{\underline{x}{\cdot}\underline{x}'}{x_\perp x'_\perp} K_1\qty(\bar{Q}x'_\perp) \underline{x}^2 \underline{x}^{\prime 2} \underline{l}^2 \underline{l}^{\prime 2} \,.
\end{equation}
Now introduce the factors $F(l_\perp)$ and $F(l'_\perp)$ giving the $l_\perp$-dependence of the target nucleons according to
\begin{equation} \label{eq:22}
    \int \dd[2]{l} F(l_\perp) l_\perp^2 = \frac{4\pi\alpha C_F\rho}{N_C^2-1} L\,xG = Q_S^2
\end{equation}
so that including $F(l_\perp) F(l'_\perp)$ in $A$ and taking $\dd[2]{l}\dd[2]{l'}$ gives
\begin{equation} \label{eq:23}
    A = \bar{Q}^2 \int \frac{\dd[2]{x}\dd[2]{x'}}{(2\pi)^4} e^{-i\underline{p}{\cdot}\qty(\underline{x}-\underline{x}')} K_1\qty(\bar{Q}x_\perp) \frac{\underline{x}{\cdot}\underline{x}'}{x_\perp x'_\perp} K_1\qty(\bar{Q}x'_\perp) T(x_\perp) T(x'_\perp)
\end{equation}
where
\begin{equation} \label{eq:24}
    T(x_\perp) = \frac{Q_S^2 x_\perp^2}{4}
\end{equation}
in this single scattering approximation.
Thus without integrating $A$ over $\dd[2]{p}$ but including the $2\alpha_{em}N_Ce_f^2$ factor gives, when $z$ is small,
\begin{equation} \label{eq:25}
    \frac{\dd{\sigma_T^{\gamma^*}}}{\dd{z} \dd[2]{b} \dd[2]{p}} = 2\alpha_{em}N_Ce_f^2 \bar{Q}^2 \int \frac{\dd[2]{x}\dd[2]{x'}}{(2\pi)^4} e^{-i\underline{p}{\cdot}\qty(\underline{x}-\underline{x}')} K_1\qty(\bar{Q}x_\perp) \frac{\underline{x}{\cdot}\underline{x}'}{x_\perp x'_\perp} K_1\qty(\bar{Q}x'_\perp) T(x_\perp) T(x'_\perp) \dd{z}
\end{equation}
agreeing with (2.14) and (2.17) of \cite{3} and also with \eqref{eq:2} and \eqref{eq:3} above.

In order to do multiple scattering it is helpful to redo the single scattering case but in a manner where the coordinate space description appears at an earlier stage, and it is to that calculation that we now turn.

\subsection{Coherent single scattering; second version}
\label{sec:3.2}

We start again with the graph of Fig.~\ref{fig:6B} and write it, including the factors $F(l_\perp)$ and $F(l'_\perp)$, as
\begin{equation} \label{eq:26}
    A = \frac{1}{8\pi^2} \int \dd[2]{l} \dd[2]{l'} \frac{F(l_\perp) F(l'_\perp) \tr{\gamma{\cdot}(p+\delta)\ \underline{\gamma}{\cdot}\underline{l}\ \gamma{\cdot}(p+l)\ \gamma{\cdot}(q-p)\ \gamma{\cdot}(p-l')\ \underline{\gamma}{\cdot}\underline{l}'}}{\qty[\qty(\underline{p}+\underline{l})^2+\bar{Q}^2] \qty[\qty(\underline{p}+\underline{l}')^2+\bar{Q}^2] 2q_+2p_+\delta_-^2}
\end{equation}
where the trace in \eqref{eq:26} is the same as in \eqref{eq:12}. Now write
\begin{equation} \label{eq:27}
    \frac{1}{\qty(\underline{p}+\underline{l})^2+\bar{Q}^2} = \int \frac{\dd[2]{x}}{2\pi} e^{i\qty(\underline{p}+\underline{l}) {\cdot} \underline{x}} K_0\qty(\bar{Q}x_\perp)
\end{equation}
so that, making sure that the $q_+$ factor in the trace survives,
\begin{multline} \label{eq:28}
    A = \frac{1}{128\pi^4q_+p_+\delta_-^2} \int \dd[2]{l} \dd[2]{l'} \dd[2]{x} \dd[2]{x'} \tr\big\{\gamma{\cdot}(p+\delta)\ \underline{\gamma}{\cdot}\underline{l}\ \qty[\gamma_+p_- - \underline{\gamma}{\cdot}\qty(\underline{p}+\underline{l})]\gamma_-q_+ \\
    \cdot \qty[\gamma_+p_- - \underline{\gamma}{\cdot}\qty(\underline{p}+\underline{l}')]\ \underline{\gamma}{\cdot}\underline{l}'
    \big\} K_0\qty(\bar{Q}x_\perp) K_0\qty(\bar{Q}x'_\perp) F(l_\perp) F(l'_\perp) e^{i\qty(\underline{p}+\underline{l}) {\cdot} \underline{x}} e^{-i\qty(\underline{p}+\underline{l}') {\cdot} \underline{x}'} \,.
\end{multline}
The $\underline{\gamma}{\cdot}\qty(\underline{p}+\underline{l})$ and $\underline{\gamma}{\cdot}\qty(\underline{p}+\underline{l}')$ terms in the trace can be written as $-i\underline{\gamma}{\cdot}\nabla_{\underline{x}}$ and $i\underline{\gamma}{\cdot}\nabla_{\underline{x}'}$, respectively, so that after integrating the $\nabla_{\underline{x}}$ and $\nabla_{\underline{x}'}$ by parts one gets
\begin{multline} \label{eq:29}
    A = \int \frac{\dd[2]{l} \dd[2]{l'} \dd[2]{x} \dd[2]{x'}}{128\pi^4p_+\delta_-^2} F(l_\perp) F(l'_\perp) \tr\big\{\gamma{\cdot}(p+\delta)\ \underline{\gamma}{\cdot}\underline{l}\ \qty[\gamma_+p_-K_0\qty(\bar{Q}x_\perp) - i\underline{\gamma}{\cdot}\nabla_{\underline{x}}K_0\qty(\bar{Q}x_\perp)] \\
    \cdot \gamma_- \qty[\gamma_+p_-K_0\qty(\bar{Q}x'_\perp) + i\underline{\gamma}{\cdot}\nabla_{\underline{x}'}K_0\qty(\bar{Q}x'_\perp)]\ \underline{\gamma}{\cdot}\underline{l}'
    \big\} e^{i\qty(\underline{p}+\underline{l}) {\cdot} \underline{x}} e^{-i\qty(\underline{p}+\underline{l}') {\cdot} \underline{x}'} \,.
\end{multline}

In order to get $\underline{l}^2$ and $\underline{l}^{\prime 2}$ factors to go along with $F(l_\perp)$ and $F(l'_\perp)$, and thus giving gluon distributions we let
\begin{equation} \label{eq:30}
    e^{i\underline{l}{\cdot}\underline{x}} \to i\underline{l}{\cdot}\underline{x} \,, \qquad e^{-i\underline{l}{\cdot}\underline{x}'} \to -i\underline{l}{\cdot}\underline{x}'
\end{equation}
and using \eqref{eq:22} one finds
\begin{multline} \label{eq:31}
    A = \qty(\frac{Q_S^2}{4})^2 \int \frac{\dd[2]{x}\dd[2]{x'} e^{i\underline{p}{\cdot}\qty(\underline{x}-\underline{x}')}}{32\pi^4p_+\delta_-^2} \tr\big\{\gamma{\cdot}(p+\delta)\qty[\underline{\gamma}{\cdot}\underline{l}\ \gamma_+p_-K_0\qty(\bar{Q}x_\perp) + ix_\perp\bar{Q} K_1\qty(\bar{Q}x_\perp)] \\
    \cdot \gamma_- \qty[\gamma_+p_-\ \underline{\gamma}{\cdot}\underline{l}' K_0\qty(\bar{Q}x'_\perp) - ix'_\perp\bar{Q} K_1\qty(\bar{Q}x'_\perp)]
    \big\} \,.
\end{multline}
Call the first term in the first $[\cdot]$ in \eqref{eq:31}, $(\underline{\gamma}{\cdot}\underline{l}\ \gamma_+p_-K_0\qty(\bar{Q}x_\perp))$, term $a$, the second term in the first $[\cdot]$ $b$, the first term in the second $[\cdot]$ $a'$, and the second term in the second $[\cdot]$ $b'$, then
\begin{equation} \label{eq:32}
    \tr_{aa'} = 8p_+p_-^2 \underline{x}{\cdot}\underline{x}' K_0\qty(\bar{Q}x_\perp) K_0\qty(\bar{Q}x'_\perp)
\end{equation}
\begin{equation} \label{eq:33}
    \tr_{bb'} = 4(p+\delta)_- x_\perp x'_\perp \bar{Q}^2 K_1\qty(\bar{Q}x_\perp) K_1\qty(\bar{Q}x'_\perp)
\end{equation}
\begin{equation} \label{eq:34}
    \tr_{ab'} = 2i \underline{p}{\cdot}\underline{x} x'_\perp \frac{\bar{Q} Q^2}{q_+} K_0\qty(\bar{Q}x_\perp) K_1\qty(\bar{Q}x'_\perp)
\end{equation}
\begin{equation} \label{eq:35}
    \tr_{ba'} = -2i \underline{p}{\cdot}\underline{x}' x_\perp \frac{\bar{Q} Q^2}{q_+} K_1\qty(\bar{Q}x_\perp) K_0\qty(\bar{Q}x'_\perp) \,.
\end{equation}

In order to evaluate the $aa'$ term, $A_{aa'}$, in \eqref{eq:31} use
\begin{equation} \label{eq:36}
    \int \dd[2]{x} e^{i\underline{p}{\cdot}\underline{x}} K_0\qty(\bar{Q}x_\perp) \underline{x}{\cdot} \int \dd[2]{x'} \underline{x}' e^{-i\underline{p}{\cdot}\underline{x}'} K_0\qty(\bar{Q}x'_\perp) = \frac{16\pi^2\underline{p}^2}{\qty[\underline{p}^2+\bar{Q}^2]^4} \,.
\end{equation}
For the $bb'$ term use
\begin{equation} \label{eq:37}
    K_1\qty(\bar{Q}x_\perp) = \frac{2\bar{Q}}{x_\perp} \pdv{\bar{Q}^2} \int \frac{\dd[2]{p}}{2\pi} \frac{e^{-i\underline{p}{\cdot}\underline{x}}}{\qty[\underline{p}^2+\bar{Q}^2]} = -\frac{2\bar{Q}}{x_\perp} \pdv{\bar{Q}^2} \int \frac{\dd[2]{p}}{2\pi} \frac{e^{-i\underline{p}{\cdot}\underline{x}}}{\qty[\underline{p}^2+\bar{Q}^2]^2}
\end{equation}
to get
\begin{equation} \label{eq:38}
    \int \dd[2]{x} e^{i\underline{p}{\cdot}\underline{x}} x_\perp K_1\qty(\bar{Q}x_\perp) \int \dd[2]{x'} e^{-i\underline{p}{\cdot}\underline{x}'} x'_\perp K_1\qty(\bar{Q}x'_\perp) = \frac{16\pi^2\bar{Q}^2}{\qty[\underline{p}^2+\bar{Q}^2]^4}
\end{equation}
and similarly for the $ab'$ term and the $ba'$ term. Each of these terms gives exactly the same contribution to $A$ so that
\begin{equation} \label{eq:39}
    A = \qty(\frac{Q_S^2}{4})^2 \frac{16}{\pi^2} \frac{\underline{p}^2\qty[\bar{Q}^2]^2}{\qty[\underline{p}^2+\bar{Q}^2]^6}
\end{equation}
which agrees with \eqref{eq:19} after inserting $F(l_\perp) F(l'_\perp)$ into the right hand side of \eqref{eq:19}, doing the $\dd[2]{l}\dd[2]{l'}$ integrations in \eqref{eq:19} and using \eqref{eq:22}.

The reason for going through this exercise of rederiving the single scattering form \eqref{eq:19} is that multiple scattering will naturally include exactly the same $\tr$ factor as in \eqref{eq:29} and this will allow us to write the final answer for multiple coherent scattering exactly as for the single scattering given in \eqref{eq:25}.

\begin{figure}[htbp]
\centering
\begin{subfigure}[b]{0.99\textwidth}
\centering
\begin{tikzpicture}
\begin{feynman} 
\vertex (p1);
\vertex [right=of p1] (l);
\vertex [below right=1cm of l] (b1);
\vertex [right=2.0cm of b1] (b2);
\vertex [right=2.0cm of b2] (b3);
\vertex [right=2.0cm of b3] (b4);
\vertex [right=1.5cm of b4] (b5);
\vertex [above right=1cm of l] (t1);
\vertex [right=2.0cm of t1] (t2);
\vertex [right=2.0cm of t2] (t3);
\vertex [right=2.0cm of t3] (t4);
\vertex [right=1.5cm of t4] (t5);
\vertex [below=1.0cm of b2] (n1tl);
\vertex [below=1.0cm of b3] (n1tr);
\vertex [below=3.0cm of b3] (n2tl);
\vertex [below=3.0cm of b4] (n2tr);
\vertex [below=0.5cm of n1tl] (n1bl);
\vertex [below=0.5cm of n1tr] (n1br);
\vertex [below=0.5cm of n2tl] (n2bl);
\vertex [below=0.5cm of n2tr] (n2br);
\vertex [below left=0.35cm of n1bl] (n1l);
\vertex [below right=0.35cm of n1br] (n1r);
\vertex [below left=0.35cm of n2bl] (n2l);
\vertex [below right=0.35cm of n2br] (n2r);
\vertex [below right=0.05cm of b1] {\(\alpha\)};
\vertex [above=0.02cm of n2tl] {\(\beta\)};

\diagram* {
  (l) -- [quarter right] (b1) -- [momentum=\(p+l_2\)] (b2) -- [momentum=\(p+l_1+l_2\)](b3) -- [fermion, momentum=\(p+l_2+\delta_1\)] (b4) -- [momentum=\(p+\delta\)] (b5),
  (l) -- [quarter left] (t1) -- (t2) -- (t3) -- [anti fermion] (t4) -- [momentum=\(q-p\)] (t5),
  (p1) -- [photon, momentum'=\(q\)] (l),
  (n1tl) -- [photon, momentum'=\(l_1\)] (b2),
  (b3) -- [photon, momentum=\(l_1-\delta_1\)] (n1tr),
  (n2tl) -- [photon, quarter left, momentum=\(l_2\)] (b1),
  (b4) -- [photon, momentum=\(l_2+\delta_1-\delta\)] (n2tr),
  (n1tl) -- (n1tr) -- [half left] (n1br) -- (n1bl) -- [half left] (n1tl),
  (n2tl) -- (n2tr) -- [half left] (n2br) -- (n2bl) -- [half left] (n2tl),
  (n1l) -- [double, double distance=1.0pt] (n1bl),
  (n1br) -- [double, double distance=1.0pt] (n1r),
  (n2l) -- [double, double distance=1.0pt] (n2bl),
  (n2br) -- [double, double distance=1.0pt] (n2r),
};

\draw[fill=black!10] (n1tl) -- (n1tr) arc (90:-90:0.25cm) -- (n1bl) arc (-90:-270:0.25cm);
\draw[fill=black!10] (n2tl) -- (n2tr) arc (90:-90:0.25cm) -- (n2bl) arc (-90:-270:0.25cm);
\end{feynman}
\end{tikzpicture}
\caption{\label{fig:7A}}
\end{subfigure}
\par\bigskip
\par\bigskip
\begin{subfigure}[b]{0.99\textwidth}
\centering
\begin{tikzpicture}
\begin{feynman} 
\vertex (p1);
\vertex [right=of p1] (l);
\vertex [below right=1cm of l] (b1);
\vertex [right=2.0cm of b1] (b2);
\vertex [right=2.0cm of b2] (b3);
\vertex [right=2.0cm of b3] (b4);
\vertex [right=1.5cm of b4] (b5);
\vertex [above right=1cm of l] (t1);
\vertex [right=2.0cm of t1] (t2);
\vertex [right=2.0cm of t2] (t3);
\vertex [right=2.0cm of t3] (t4);
\vertex [right=1.5cm of t4] (t5);
\vertex [below=1.0cm of b1] (n1tl);
\vertex [below=1.0cm of b3] (n1tr);
\vertex [below=1.5cm of b2] (n2x);
\vertex [below=1.5cm of n2x] (n2tl);
\vertex [below=3.0cm of b4] (n2tr);
\vertex [below=0.5cm of n1tl] (n1bl);
\vertex [below=0.5cm of n1tr] (n1br);
\vertex [below=0.5cm of n2tl] (n2bl);
\vertex [below=0.5cm of n2tr] (n2br);
\vertex [below left=0.35cm of n1bl] (n1l);
\vertex [below right=0.35cm of n1br] (n1r);
\vertex [below left=0.35cm of n2bl] (n2l);
\vertex [below right=0.35cm of n2br] (n2r);
\vertex [below right=0.05cm of b2] {\(\alpha\)};
\vertex [above left=0.05cm of n2tl] {\(\beta\)};

\diagram* {
  (l) -- [quarter right] (b1) -- [momentum=\(p+l_1\)] (b2) -- [momentum=\(p+l_1+l_2\)](b3) -- [fermion, momentum=\(p+l_2+\delta_1\)] (b4) -- [momentum=\(p+\delta\)] (b5),
  (l) -- [quarter left] (t1) -- (t2) -- (t3) -- [anti fermion] (t4) -- [momentum=\(q-p\)] (t5),
  (p1) -- [photon, momentum'=\(q\)] (l),
  (n1tl) -- [photon, momentum'=\(l_1\)] (b1),
  (b3) -- [photon, momentum=\(l_1-\delta_1\)] (n1tr),
  (n2tl) -- [photon, momentum'=\(l_2\)] (n2x) -- [photon] (b2),
  (b4) -- [photon, momentum=\(l_2+\delta_1-\delta\)] (n2tr),
  (n1tl) -- (n1tr) -- [half left] (n1br) -- (n1bl) -- [half left] (n1tl),
  (n2tl) -- (n2tr) -- [half left] (n2br) -- (n2bl) -- [half left] (n2tl),
  (n1l) -- [double, double distance=1.0pt] (n1bl),
  (n1br) -- [double, double distance=1.0pt] (n1r),
  (n2l) -- [double, double distance=1.0pt] (n2bl),
  (n2br) -- [double, double distance=1.0pt] (n2r),
};

\draw[fill=black!10] (n1tl) -- (n1tr) arc (90:-90:0.25cm) -- (n1bl) arc (-90:-270:0.25cm);
\draw[fill=black!10] (n2tl) -- (n2tr) arc (90:-90:0.25cm) -- (n2bl) arc (-90:-270:0.25cm);

\diagram* {
  (n2x) -- [photon] (b2)
};
\end{feynman}
\end{tikzpicture}
\caption{\label{fig:7B}}
\end{subfigure}
\par\bigskip
\par\bigskip
\begin{subfigure}[b]{0.99\textwidth}
\centering
\begin{tikzpicture}
\begin{feynman} 
\vertex (p1);
\vertex [right=of p1] (l);
\vertex [below right=1cm of l] (b1);
\vertex [right=2.0cm of b1] (b2);
\vertex [right=2.0cm of b2] (b3);
\vertex [right=2.0cm of b3] (b4);
\vertex [right=1.5cm of b4] (b5);
\vertex [above right=1cm of l] (t1);
\vertex [right=2.0cm of t1] (t2);
\vertex [right=2.0cm of t2] (t3);
\vertex [right=2.0cm of t3] (t4);
\vertex [right=1.5cm of t4] (t5);
\vertex [below=1.5cm of b1] (n1tl);
\vertex [below=1.5cm of b2] (n1tr);
\vertex [below=1.5cm of b3] (n2tl);
\vertex [below=1.5cm of b4] (n2tr);
\vertex [below=0.5cm of n1tl] (n1bl);
\vertex [below=0.5cm of n1tr] (n1br);
\vertex [below=0.5cm of n2tl] (n2bl);
\vertex [below=0.5cm of n2tr] (n2br);
\vertex [below left=0.35cm of n1bl] (n1l);
\vertex [below right=0.35cm of n1br] (n1r);
\vertex [below left=0.35cm of n2bl] (n2l);
\vertex [below right=0.35cm of n2br] (n2r);
\vertex [below left=0.05cm of b3] {\(\alpha\)};
\vertex [above left=0.05cm of n2tl] {\(\beta\)};

\diagram* {
  (l) -- [quarter right] (b1) -- [momentum=\(p+l_1\)] (b2) -- [momentum=\(p+l_1+\delta_1\)](b3) -- [fermion, momentum=\(p+l_2+\delta_1\)] (b4) -- [momentum=\(p+\delta\)] (b5),
  (l) -- [quarter left] (t1) -- (t2) -- (t3) -- [anti fermion] (t4) -- [momentum=\(q-p\)] (t5),
  (p1) -- [photon, momentum'=\(q\)] (l),
  (n1tl) -- [photon, momentum'=\(l_1\)] (b1),
  (b2) -- [photon, momentum=\(l_1-\delta_1\)] (n1tr),
  (n2tl) -- [photon, momentum'=\(l_2\)] (b3),
  (b4) -- [photon, momentum=\(l_2+\delta_1-\delta\)] (n2tr),
  (n1tl) -- (n1tr) -- [half left] (n1br) -- (n1bl) -- [half left] (n1tl),
  (n2tl) -- (n2tr) -- [half left] (n2br) -- (n2bl) -- [half left] (n2tl),
  (n1l) -- [double, double distance=1.0pt] (n1bl),
  (n1br) -- [double, double distance=1.0pt] (n1r),
  (n2l) -- [double, double distance=1.0pt] (n2bl),
  (n2br) -- [double, double distance=1.0pt] (n2r),
};

\draw[fill=black!10] (n1tl) -- (n1tr) arc (90:-90:0.25cm) -- (n1bl) arc (-90:-270:0.25cm);
\draw[fill=black!10] (n2tl) -- (n2tr) arc (90:-90:0.25cm) -- (n2bl) arc (-90:-270:0.25cm);
\end{feynman}
\end{tikzpicture}
\caption{\label{fig:7C}}
\end{subfigure}
\caption{An illustration of two dipole scatterings with nucleons in the nucleus. A, B and C show the three types of scatterings. \label{fig:7}}
\end{figure}
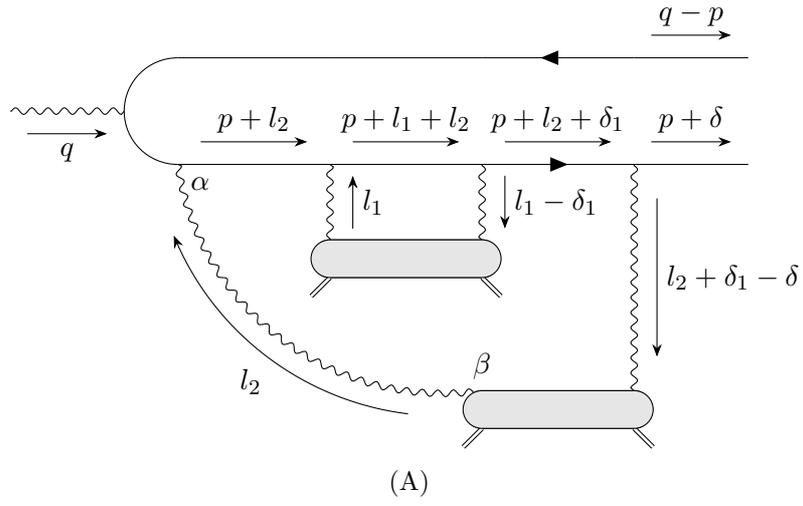
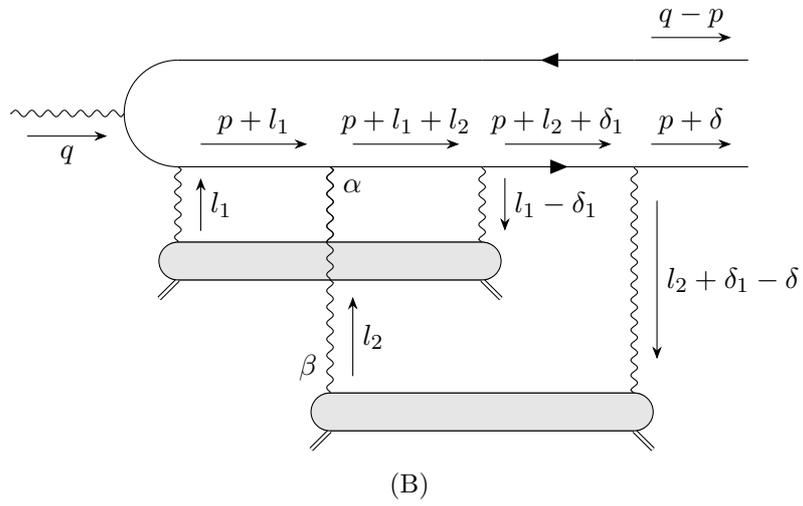
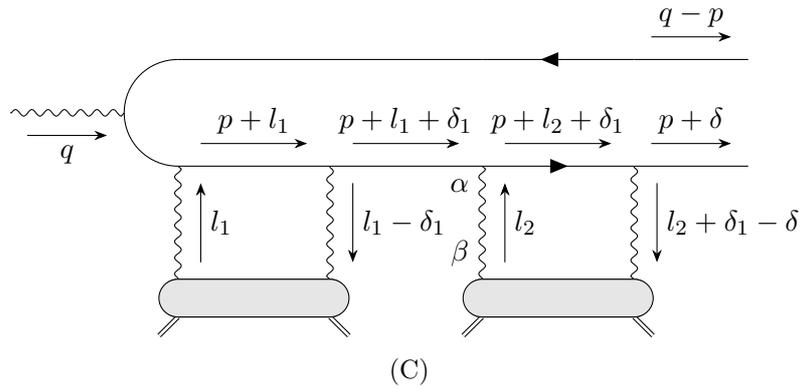

\subsection{Coherent multiple scattering}
\label{sec:3.3}

\begin{figure}[htbp]
\centering
\begin{tikzpicture}
\begin{feynman} 
\vertex (p1);
\vertex [right=of p1] (l);
\vertex [below right=1cm of l] (b1);
\vertex [right=1.5cm of b1] (b2);
\vertex [right=2.0cm of b2] (b3);
\vertex [right=1.0cm of b3] (b4);
\vertex [right=2.0cm of b4] (b5);
\vertex [right=1.5cm of b5] (b6);
\vertex [above right=1cm of l] (t1);
\vertex [right=1.5cm of t1] (t2);
\vertex [right=2.0cm of t2] (t3);
\vertex [right=1.0cm of t3] (t4);
\vertex [right=2.0cm of t4] (t5);
\vertex [right=1.5cm of t5] (t6);
\vertex [below=1.5cm of b2] (n1tl);
\vertex [below=1.5cm of b3] (n1tr);
\vertex [below=3.5cm of b4] (n2tl);
\vertex [below=3.5cm of b5] (n2tr);
\vertex [below=0.5cm of n1tl] (n1bl);
\vertex [below=0.5cm of n1tr] (n1br);
\vertex [below=0.5cm of n2tl] (n2bl);
\vertex [below=0.5cm of n2tr] (n2br);
\vertex [below left=0.35cm of n1bl] (n1l);
\vertex [below right=0.35cm of n1br] (n1r);
\vertex [below left=0.35cm of n2bl] (n2l);
\vertex [below right=0.35cm of n2br] (n2r);

\diagram* {
  (l) -- [quarter right] (b1) -- (b2) -- [momentum=\(p+l_1+l_2\)] (b3) -- [fermion] (b4) -- (b5) -- [momentum=\(p+\delta\)] (b6),
  (l) -- [quarter left] (t1) -- (t2) -- (t3) -- [anti fermion] (t4) -- (t5) -- [momentum=\(q-p\)] (t6),
  (p1) -- [photon, momentum'=\(q\)] (l),
  (n1tl) -- [scalar, out=180, in=-115, momentum'=\(l_1\)] (l),
  (b5) -- [scalar, out=-135, in=0, momentum'=\(l_1-\delta_1\)] (n1tr),
  (n2tl) -- [scalar, out=180, in=-115, momentum=\(l_2\)] (l),
  (b5) -- [photon, momentum=\(l_2+\delta_1-\delta\)] (n2tr),
  (n1tl) -- (n1tr) -- [half left] (n1br) -- (n1bl) -- [half left] (n1tl),
  (n2tl) -- (n2tr) -- [half left] (n2br) -- (n2bl) -- [half left] (n2tl),
  (n1l) -- [double, double distance=1.0pt] (n1bl),
  (n1br) -- [double, double distance=1.0pt] (n1r),
  (n2l) -- [double, double distance=1.0pt] (n2bl),
  (n2br) -- [double, double distance=1.0pt] (n2r),
};

\draw[fill=black!10] (n1tl) -- (n1tr) arc (90:-90:0.25cm) -- (n1bl) arc (-90:-270:0.25cm);
\draw[fill=black!10] (n2tl) -- (n2tr) arc (90:-90:0.25cm) -- (n2bl) arc (-90:-270:0.25cm);
\end{feynman}
\end{tikzpicture}
\caption{The result of using the Ward identities in the graphs of Fig.~\ref{fig:7}. \label{fig:8}}
\end{figure}
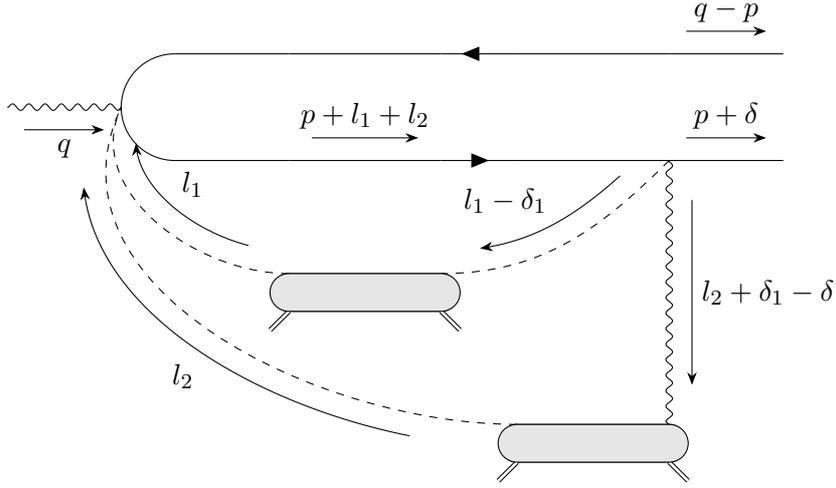

In this section we study two coherent scatterings in the amplitude in detail. After this calculation is completed it will become apparent how to do the multiple coherent scattering case with little further work. Consider the three graphs in Fig.~\ref{fig:7}. We are going to show that these three graphs can be reduced to the single graph shown in Fig.~\ref{fig:8}.

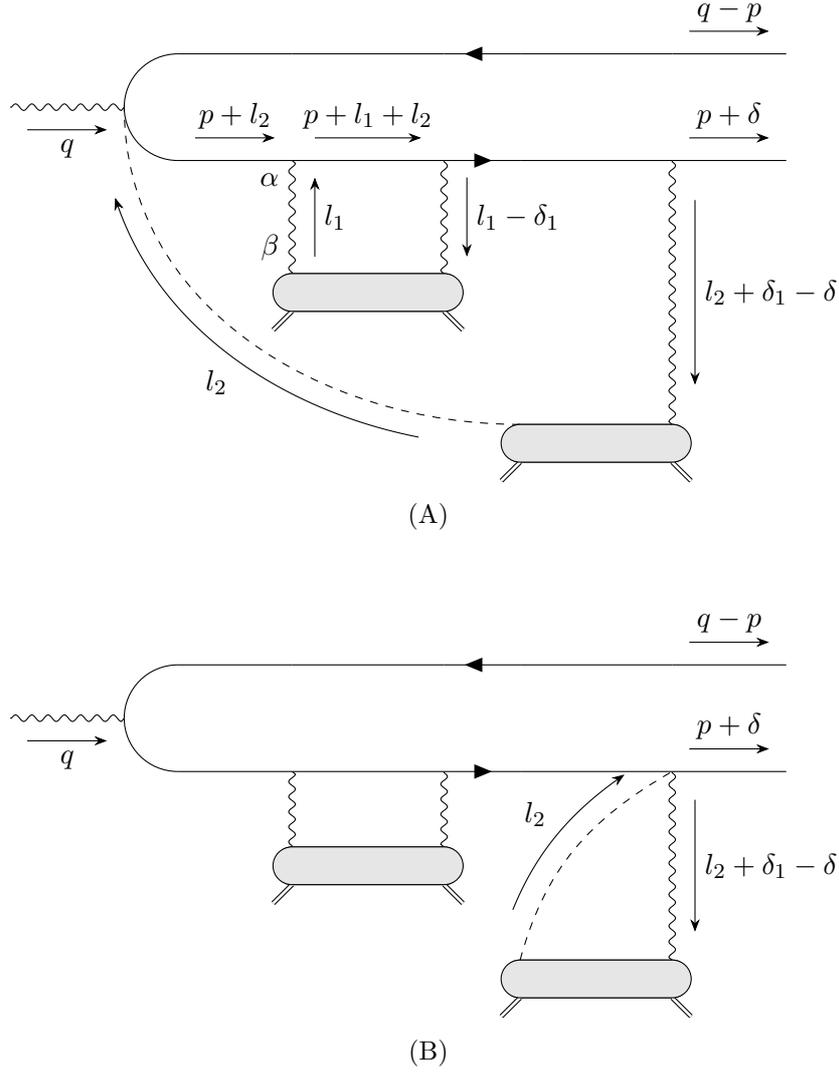
\begin{figure}[htbp]
\centering
\begin{subfigure}[b]{0.99\textwidth}
\centering
\begin{tikzpicture}
\begin{feynman} 
\vertex (p1);
\vertex [right=of p1] (l);
\vertex [below right=1cm of l] (b1);
\vertex [right=1.5cm of b1] (b2);
\vertex [right=2.0cm of b2] (b3);
\vertex [right=1.0cm of b3] (b4);
\vertex [right=2.0cm of b4] (b5);
\vertex [right=1.5cm of b5] (b6);
\vertex [above right=1cm of l] (t1);
\vertex [right=1.5cm of t1] (t2);
\vertex [right=2.0cm of t2] (t3);
\vertex [right=1.0cm of t3] (t4);
\vertex [right=2.0cm of t4] (t5);
\vertex [right=1.5cm of t5] (t6);
\vertex [below=1.5cm of b2] (n1tl);
\vertex [below=1.5cm of b3] (n1tr);
\vertex [below=3.5cm of b4] (n2tl);
\vertex [below=3.5cm of b5] (n2tr);
\vertex [below=0.5cm of n1tl] (n1bl);
\vertex [below=0.5cm of n1tr] (n1br);
\vertex [below=0.5cm of n2tl] (n2bl);
\vertex [below=0.5cm of n2tr] (n2br);
\vertex [below left=0.35cm of n1bl] (n1l);
\vertex [below right=0.35cm of n1br] (n1r);
\vertex [below left=0.35cm of n2bl] (n2l);
\vertex [below right=0.35cm of n2br] (n2r);
\vertex [below left=0.05cm of b2] {\(\alpha\)};
\vertex [above left=0.05cm of n1tl] {\(\beta\)};

\diagram* {
  (l) -- [quarter right] (b1) -- [momentum=\(p+l_2\)] (b2) -- [momentum=\(p+l_1+l_2\)] (b3) -- [fermion] (b4) -- (b5) -- [momentum=\(p+\delta\)] (b6),
  (l) -- [quarter left] (t1) -- (t2) -- (t3) -- [anti fermion] (t4) -- (t5) -- [momentum=\(q-p\)] (t6),
  (p1) -- [photon, momentum'=\(q\)] (l),
  (n1tl) -- [photon, momentum'=\(l_1\)] (b2),
  (b3) -- [photon, momentum=\(l_1-\delta_1\)] (n1tr),
  (n2tl) -- [scalar, out=180, in=-90, momentum=\(l_2\)] (l),
  (b5) -- [photon, momentum=\(l_2+\delta_1-\delta\)] (n2tr),
  (n1tl) -- (n1tr) -- [half left] (n1br) -- (n1bl) -- [half left] (n1tl),
  (n2tl) -- (n2tr) -- [half left] (n2br) -- (n2bl) -- [half left] (n2tl),
  (n1l) -- [double, double distance=1.0pt] (n1bl),
  (n1br) -- [double, double distance=1.0pt] (n1r),
  (n2l) -- [double, double distance=1.0pt] (n2bl),
  (n2br) -- [double, double distance=1.0pt] (n2r),
};

\draw[fill=black!10] (n1tl) -- (n1tr) arc (90:-90:0.25cm) -- (n1bl) arc (-90:-270:0.25cm);
\draw[fill=black!10] (n2tl) -- (n2tr) arc (90:-90:0.25cm) -- (n2bl) arc (-90:-270:0.25cm);
\end{feynman}
\end{tikzpicture}
\caption{\label{fig:9A}}
\end{subfigure}
\par\bigskip
\par\bigskip
\begin{subfigure}[b]{0.99\textwidth}
\centering
\begin{tikzpicture}
\begin{feynman} 
\vertex (p1);
\vertex [right=of p1] (l);
\vertex [below right=1cm of l] (b1);
\vertex [right=1.5cm of b1] (b2);
\vertex [right=2.0cm of b2] (b3);
\vertex [right=1.0cm of b3] (b4);
\vertex [right=2.0cm of b4] (b5);
\vertex [right=1.5cm of b5] (b6);
\vertex [above right=1cm of l] (t1);
\vertex [right=1.5cm of t1] (t2);
\vertex [right=2.0cm of t2] (t3);
\vertex [right=1.0cm of t3] (t4);
\vertex [right=2.0cm of t4] (t5);
\vertex [right=1.5cm of t5] (t6);
\vertex [below=1.0cm of b2] (n1tl);
\vertex [below=1.0cm of b3] (n1tr);
\vertex [below=2.5cm of b4] (n2tl);
\vertex [below=2.5cm of b5] (n2tr);
\vertex [below=0.5cm of n1tl] (n1bl);
\vertex [below=0.5cm of n1tr] (n1br);
\vertex [below=0.5cm of n2tl] (n2bl);
\vertex [below=0.5cm of n2tr] (n2br);
\vertex [below left=0.35cm of n1bl] (n1l);
\vertex [below right=0.35cm of n1br] (n1r);
\vertex [below left=0.35cm of n2bl] (n2l);
\vertex [below right=0.35cm of n2br] (n2r);

\diagram* {
  (l) -- [quarter right] (b1) -- (b2) -- (b3) -- [fermion] (b4) -- (b5) -- [momentum=\(p+\delta\)] (b6),
  (l) -- [quarter left] (t1) -- (t2) -- (t3) -- [anti fermion] (t4) -- (t5) -- [momentum=\(q-p\)] (t6),
  (p1) -- [photon, momentum'=\(q\)] (l),
  (n1tl) -- [photon] (b2),
  (b3) -- [photon] (n1tr),
  (n2tl) -- [scalar, out=75, in=-150, momentum=\(l_2\)] (b5),
  (b5) -- [photon, momentum=\(l_2+\delta_1-\delta\)] (n2tr),
  (n1tl) -- (n1tr) -- [half left] (n1br) -- (n1bl) -- [half left] (n1tl),
  (n2tl) -- (n2tr) -- [half left] (n2br) -- (n2bl) -- [half left] (n2tl),
  (n1l) -- [double, double distance=1.0pt] (n1bl),
  (n1br) -- [double, double distance=1.0pt] (n1r),
  (n2l) -- [double, double distance=1.0pt] (n2bl),
  (n2br) -- [double, double distance=1.0pt] (n2r),
};

\draw[fill=black!10] (n1tl) -- (n1tr) arc (90:-90:0.25cm) -- (n1bl) arc (-90:-270:0.25cm);
\draw[fill=black!10] (n2tl) -- (n2tr) arc (90:-90:0.25cm) -- (n2bl) arc (-90:-270:0.25cm);
\end{feynman}
\end{tikzpicture}
\caption{\label{fig:9B}}
\end{subfigure}
\caption{An intermediate step in going from the graphs of Fig.~\ref{fig:7} to that of Fig.~\ref{fig:8}. \label{fig:9}}
\end{figure}

We begin our evaluation of the graphs in Fig.~\ref{fig:7} by doing the $\dd{l_{2-}}$ integration. In each of the graphs in Fig.~\ref{fig:7} the only singularity in the upper half $l_{2-}$-plane comes from the $\frac{1}{l_{2-}-i\epsilon}$ light cone gauge singularity (corresponding to the $\frac{\eta_\beta l_{2\alpha}}{l_{2-}-i\epsilon}$ term in the $G_{\beta\alpha}(l_2)$ propagator). Rearranging the propagators of the fermion line using the $l_2{\cdot}\gamma$ term in the numerator, as in \eqref{eq:11}, one finds the result shown in Fig.~\ref{fig:9}. (For simplicity the three gluon coupling terms in the Ward identities are not shown explicitly.) The graph of Fig.~\ref{fig:9B} does not have a $\frac{\dd{\underline{l}_2^2}}{\underline{l}_2^2}$ term necessary to get the gluon distribution of the nucleon so we ignore this graph and only keep the graph of Fig.~\ref{fig:9A}.

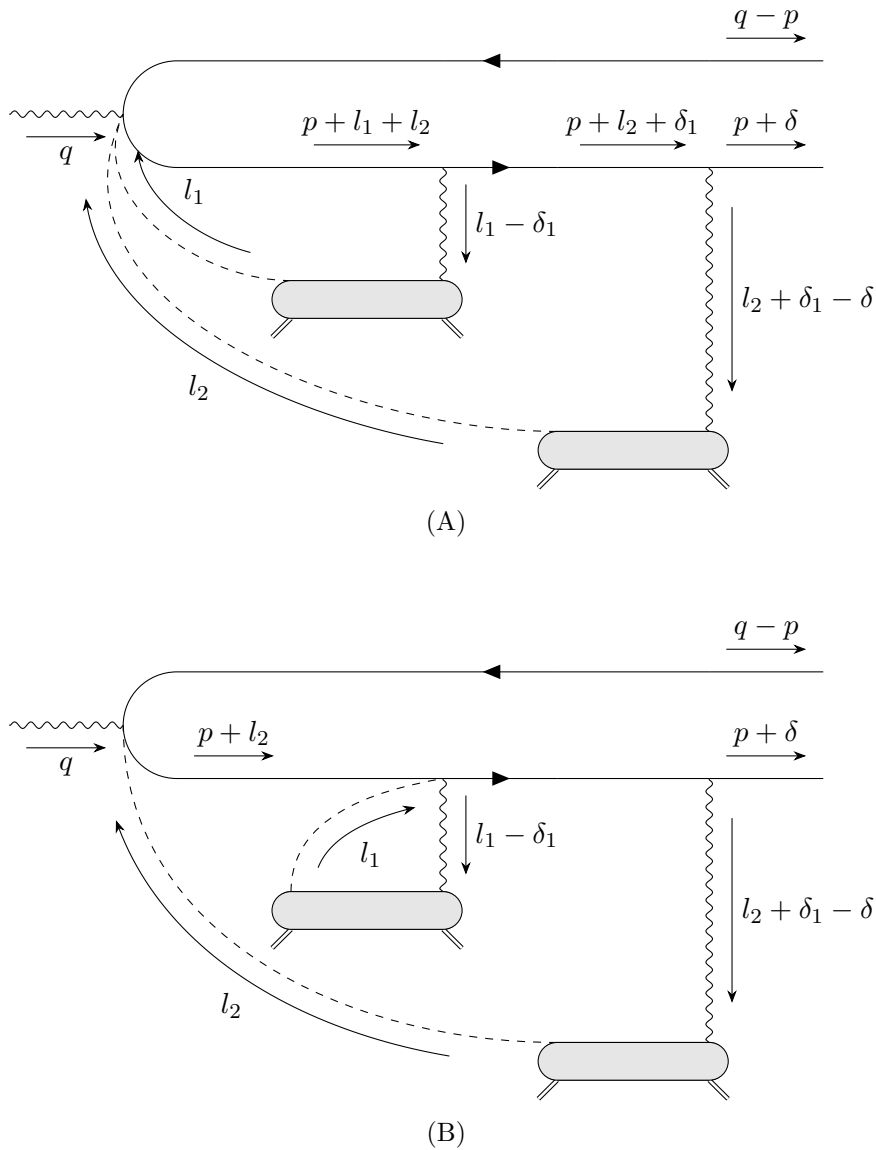
\begin{figure}[htbp]
\centering
\begin{subfigure}[b]{0.99\textwidth}
\centering
\begin{tikzpicture}
\begin{feynman} 
\vertex (p1);
\vertex [right=of p1] (l);
\vertex [below right=1cm of l] (b1);
\vertex [right=1.5cm of b1] (b2);
\vertex [right=2.0cm of b2] (b3);
\vertex [right=1.5cm of b3] (b4);
\vertex [right=2.0cm of b4] (b5);
\vertex [right=1.5cm of b5] (b6);
\vertex [above right=1cm of l] (t1);
\vertex [right=1.5cm of t1] (t2);
\vertex [right=2.0cm of t2] (t3);
\vertex [right=1.5cm of t3] (t4);
\vertex [right=2.0cm of t4] (t5);
\vertex [right=1.5cm of t5] (t6);
\vertex [below=1.5cm of b2] (n1tl);
\vertex [below=1.5cm of b3] (n1tr);
\vertex [below=3.5cm of b4] (n2tl);
\vertex [below=3.5cm of b5] (n2tr);
\vertex [below=0.5cm of n1tl] (n1bl);
\vertex [below=0.5cm of n1tr] (n1br);
\vertex [below=0.5cm of n2tl] (n2bl);
\vertex [below=0.5cm of n2tr] (n2br);
\vertex [below left=0.35cm of n1bl] (n1l);
\vertex [below right=0.35cm of n1br] (n1r);
\vertex [below left=0.35cm of n2bl] (n2l);
\vertex [below right=0.35cm of n2br] (n2r);

\diagram* {
  (l) -- [quarter right] (b1) -- (b2) -- [momentum=\(p+l_1+l_2\)] (b3) -- [fermion] (b4) -- [momentum=\(p+l_2+\delta_1\)] (b5) -- [momentum=\(p+\delta\)] (b6),
  (l) -- [quarter left] (t1) -- (t2) -- (t3) -- [anti fermion] (t4) -- (t5) -- [momentum=\(q-p\)] (t6),
  (p1) -- [photon, momentum'=\(q\)] (l),
  (n1tl) -- [scalar, out=180, in=-115, momentum'=\(l_1\)] (l),
  (b3) -- [photon, momentum=\(l_1-\delta_1\)] (n1tr),
  (n2tl) -- [scalar, out=180, in=-115, momentum=\(l_2\)] (l),
  (b5) -- [photon, momentum=\(l_2+\delta_1-\delta\)] (n2tr),
  (n1tl) -- (n1tr) -- [half left] (n1br) -- (n1bl) -- [half left] (n1tl),
  (n2tl) -- (n2tr) -- [half left] (n2br) -- (n2bl) -- [half left] (n2tl),
  (n1l) -- [double, double distance=1.0pt] (n1bl),
  (n1br) -- [double, double distance=1.0pt] (n1r),
  (n2l) -- [double, double distance=1.0pt] (n2bl),
  (n2br) -- [double, double distance=1.0pt] (n2r),
};

\draw[fill=black!10] (n1tl) -- (n1tr) arc (90:-90:0.25cm) -- (n1bl) arc (-90:-270:0.25cm);
\draw[fill=black!10] (n2tl) -- (n2tr) arc (90:-90:0.25cm) -- (n2bl) arc (-90:-270:0.25cm);
\end{feynman}
\end{tikzpicture}
\caption{\label{fig:10A}}
\end{subfigure}
\par\bigskip
\par\bigskip
\begin{subfigure}[b]{0.99\textwidth}
\centering
\begin{tikzpicture}
\begin{feynman} 
\vertex (p1);
\vertex [right=of p1] (l);
\vertex [below right=1cm of l] (b1);
\vertex [right=1.5cm of b1] (b2);
\vertex [right=2.0cm of b2] (b3);
\vertex [right=1.5cm of b3] (b4);
\vertex [right=2.0cm of b4] (b5);
\vertex [right=1.5cm of b5] (b6);
\vertex [above right=1cm of l] (t1);
\vertex [right=1.5cm of t1] (t2);
\vertex [right=2.0cm of t2] (t3);
\vertex [right=1.5cm of t3] (t4);
\vertex [right=2.0cm of t4] (t5);
\vertex [right=1.5cm of t5] (t6);
\vertex [below=1.5cm of b2] (n1tl);
\vertex [below=1.5cm of b3] (n1tr);
\vertex [below=3.5cm of b4] (n2tl);
\vertex [below=3.5cm of b5] (n2tr);
\vertex [below=0.5cm of n1tl] (n1bl);
\vertex [below=0.5cm of n1tr] (n1br);
\vertex [below=0.5cm of n2tl] (n2bl);
\vertex [below=0.5cm of n2tr] (n2br);
\vertex [below left=0.35cm of n1bl] (n1l);
\vertex [below right=0.35cm of n1br] (n1r);
\vertex [below left=0.35cm of n2bl] (n2l);
\vertex [below right=0.35cm of n2br] (n2r);

\diagram* {
  (l) -- [quarter right] (b1) -- [momentum=\(p+l_2\)] (b2) -- (b3) -- [fermion] (b4) -- (b5) -- [momentum=\(p+\delta\)] (b6),
  (l) -- [quarter left] (t1) -- (t2) -- (t3) -- [anti fermion] (t4) -- (t5) -- [momentum=\(q-p\)] (t6),
  (p1) -- [photon, momentum'=\(q\)] (l),
  (n1tl) -- [scalar, out=90, in=-170, momentum'=\(l_1\)] (b3),
  (b3) -- [photon, momentum=\(l_1-\delta_1\)] (n1tr),
  (n2tl) -- [scalar, out=180, in=-90, momentum=\(l_2\)] (l),
  (b5) -- [photon, momentum=\(l_2+\delta_1-\delta\)] (n2tr),
  (n1tl) -- (n1tr) -- [half left] (n1br) -- (n1bl) -- [half left] (n1tl),
  (n2tl) -- (n2tr) -- [half left] (n2br) -- (n2bl) -- [half left] (n2tl),
  (n1l) -- [double, double distance=1.0pt] (n1bl),
  (n1br) -- [double, double distance=1.0pt] (n1r),
  (n2l) -- [double, double distance=1.0pt] (n2bl),
  (n2br) -- [double, double distance=1.0pt] (n2r),
};

\draw[fill=black!10] (n1tl) -- (n1tr) arc (90:-90:0.25cm) -- (n1bl) arc (-90:-270:0.25cm);
\draw[fill=black!10] (n2tl) -- (n2tr) arc (90:-90:0.25cm) -- (n2bl) arc (-90:-270:0.25cm);
\end{feynman}
\end{tikzpicture}
\caption{\label{fig:10B}}
\end{subfigure}
\caption{Further steps in using the Ward identities to go from the graphs of Fig.~\ref{fig:7} to that of Fig.~\ref{fig:8}. \label{fig:10}}
\end{figure}

Next consider the $\dd{l_{1-}}$ integration of the graph in Fig.~\ref{fig:9A}. Again the only singularity in the upper half $l_{1-}$-plane comes from light cone gauge singularity. Again applying the Ward identities coming from the $l_1{\cdot}\gamma$ vertex the graph of Fig.~\ref{fig:9A} becomes that of Fig.~\ref{fig:10}. Again only the graph of Fig.~\ref{fig:10A} gives an important term. Finally doing the integral over the singularity $\frac{1}{\delta_{1-}-i\epsilon}$, from the $(l_1-\delta)$-line, in the $\dd{\delta_{1-}}$ integration and applying the Ward identity one ends up with the graph shown in Fig.~\ref{fig:8}. The dashed lines in Fig.~\ref{fig:8} carry only transverse momentum and color charge while the attachments of the dashed lines to the $p$-line carry no vector indices.

\begin{figure}[htbp]
\centering
\begin{tikzpicture}
\begin{feynman} 
\vertex (p1);
\vertex [right=of p1] (l);
\vertex [below right=0.707cm of l] (b1);
\vertex [right=0.5cm of b1] (b2);
\vertex [right=1.0cm of b2] (b3);
\vertex [right=1.0cm of b3] (b4);
\vertex [right=0.5cm of b4] (b5);
\vertex [right=1.5cm of b5] (b6);
\vertex [right=1.5cm of b6] (b7);
\vertex [right=0.5cm of b7] (b8);
\vertex [right=1.0cm of b8] (b9);
\vertex [right=1.0cm of b9] (b10);
\vertex [right=0.5cm of b10] (b11);
\vertex [above right=0.707cm of b11] (r);
\vertex [right=of r] (p2);
\vertex [above right=0.707cm of l] (t1);
\vertex [right=0.5cm of t1] (t2);
\vertex [right=1.0cm of t2] (t3);
\vertex [right=1.0cm of t3] (t4);
\vertex [right=0.5cm of t4] (t5);
\vertex [right=1.5cm of t5] (t6);
\vertex [right=1.5cm of t6] (t7);
\vertex [right=0.5cm of t7] (t8);
\vertex [right=1.0cm of t8] (t9);
\vertex [right=1.0cm of t9] (t10);
\vertex [right=0.5cm of t10] (t11);
\vertex [above=0.5cm of t6] (tt);
\vertex [below=0.5cm of b6] (bb);
\vertex [below=0.5cm of b2] (n1tl);
\vertex [below=0.5cm of b3] (n1tr);
\vertex [below=1.5cm of b2] (n2tl);
\vertex [below=1.5cm of b3] (n2tr);
\vertex [below=3.0cm of b3] (n3tl);
\vertex [below=3.0cm of b4] (n3tr);
\vertex [below=0.5cm of b9] (n4tl);
\vertex [below=0.5cm of b10] (n4tr);
\vertex [below=1.5cm of b9] (n5tl);
\vertex [below=1.5cm of b10] (n5tr);
\vertex [below=3.0cm of b8] (n6tl);
\vertex [below=3.0cm of b9] (n6tr);
\vertex [below=0.5cm of n1tl] (n1bl);
\vertex [below=0.5cm of n1tr] (n1br);
\vertex [below=0.5cm of n2tl] (n2bl);
\vertex [below=0.5cm of n2tr] (n2br);
\vertex [below=0.5cm of n3tl] (n3bl);
\vertex [below=0.5cm of n3tr] (n3br);
\vertex [below=0.5cm of n4tl] (n4bl);
\vertex [below=0.5cm of n4tr] (n4br);
\vertex [below=0.5cm of n5tl] (n5bl);
\vertex [below=0.5cm of n5tr] (n5br);
\vertex [below=0.5cm of n6tl] (n6bl);
\vertex [below=0.5cm of n6tr] (n6br);
\vertex [below left=0.35cm of n1bl] (n1l);
\vertex [below right=0.35cm of n1br] (n1r);
\vertex [below left=0.35cm of n2bl] (n2l);
\vertex [below right=0.35cm of n2br] (n2r);
\vertex [below left=0.35cm of n3bl] (n3l);
\vertex [below right=0.35cm of n3br] (n3r);
\vertex [below left=0.35cm of n4bl] (n4l);
\vertex [below right=0.35cm of n4br] (n4r);
\vertex [below left=0.35cm of n5bl] (n5l);
\vertex [below right=0.35cm of n5br] (n5r);
\vertex [below left=0.35cm of n6bl] (n6l);
\vertex [below right=0.35cm of n6br] (n6r);
\vertex [right=0.5cm of n1bl] (n1b);
\vertex [right=0.5cm of n1tl] (n1t);
\vertex [right=0.5cm of n2bl] (n2b);
\vertex [right=0.5cm of n2tl] (n2t);
\vertex [below=0.12cm of n1b] (n12t);
\vertex [above=0.12cm of n2t] (n12b);
\vertex [right=0.5cm of n4bl] (n4b);
\vertex [right=0.5cm of n4tl] (n4t);
\vertex [right=0.5cm of n5bl] (n5b);
\vertex [right=0.5cm of n5tl] (n5t);
\vertex [below=0.12cm of n4b] (n45t);
\vertex [above=0.12cm of n5t] (n45b);

\diagram* {
  (l) -- [quarter right] (b1) -- [momentum=\(p+l_1+\ldots+l_{N+1}\)] (b4) -- (b5) -- [momentum'=\(p+\delta\)] (b6) -- [fermion] (b7) -- [momentum=\(p+l'_1+\ldots+l'_{M+1}\)] (b11) -- [quarter right] (r),
  (l) -- [quarter left] (t1) -- (t2) -- (t3) -- (t4) -- (t5) -- [anti fermion, momentum=\(q-p\)] (t6) -- (t7) -- (t8) -- (t9) -- (t10) -- (t11) -- [quarter left] (r),
  (p1) -- [photon, momentum'=\(q\)] (l),
  (r) -- [photon, momentum'=\(q\)] (p2),
  (n1tl) -- [scalar, out=160, in=-105] (l),
  (b4) -- [scalar, out=-135, in=20] (n1tr),
  (n2tl) -- [scalar, out=160, in=-110] (l),
  (b4) -- [scalar,out=-115, in=20] (n2tr),
  (n3tl) -- [scalar, out=180, in=-115, momentum=\(l_{N+1}\)] (l),
  (b4) -- [photon, momentum=\(l_{N+1}+\delta_N-\delta\)] (n3tr),
  (n4tl) -- [scalar, out=160, in=-45] (b8),
  (r) -- [scalar, out=-75, in=20] (n4tr),
  (n5tl) -- [scalar, out=160, in=-65] (b8),
  (r) -- [scalar, out=-70, in=20] (n5tr),
  (n6tl) -- [photon] (b8),
  (r) -- [scalar, out=-65, in=0, momentum=\(l'_{M+1}\)] (n6tr),
  (n1tl) -- (n1tr) -- [half left] (n1br) -- (n1bl) -- [half left] (n1tl),
  (n2tl) -- (n2tr) -- [half left] (n2br) -- (n2bl) -- [half left] (n2tl),
  (n3tl) -- (n3tr) -- [half left] (n3br) -- (n3bl) -- [half left] (n3tl),
  (n4tl) -- (n4tr) -- [half left] (n4br) -- (n4bl) -- [half left] (n4tl),
  (n5tl) -- (n5tr) -- [half left] (n5br) -- (n5bl) -- [half left] (n5tl),
  (n6tl) -- (n6tr) -- [half left] (n6br) -- (n6bl) -- [half left] (n6tl),
  (n1l) -- [double, double distance=1.0pt] (n1bl),
  (n1br) -- [double, double distance=1.0pt] (n1r),
  (n2l) -- [double, double distance=1.0pt] (n2bl),
  (n2br) -- [double, double distance=1.0pt] (n2r),
  (n3l) -- [double, double distance=1.0pt] (n3bl),
  (n3br) -- [double, double distance=1.0pt] (n3r),
  (n4l) -- [double, double distance=1.0pt] (n4bl),
  (n4br) -- [double, double distance=1.0pt] (n4r),
  (n5l) -- [double, double distance=1.0pt] (n5bl),
  (n5br) -- [double, double distance=1.0pt] (n5r),
  (n6l) -- [double, double distance=1.0pt] (n6bl),
  (n6br) -- [double, double distance=1.0pt] (n6r),
  (n12t) -- [ghost] (n12b),
  (n45t) -- [ghost] (n45b),
};

\draw[fill=black!10] (n1tl) -- (n1tr) arc (90:-90:0.25cm) -- (n1bl) arc (-90:-270:0.25cm);
\draw[fill=black!10] (n2tl) -- (n2tr) arc (90:-90:0.25cm) -- (n2bl) arc (-90:-270:0.25cm);
\draw[fill=black!10] (n3tl) -- (n3tr) arc (90:-90:0.25cm) -- (n3bl) arc (-90:-270:0.25cm);
\draw[fill=black!10] (n4tl) -- (n4tr) arc (90:-90:0.25cm) -- (n4bl) arc (-90:-270:0.25cm);
\draw[fill=black!10] (n5tl) -- (n5tr) arc (90:-90:0.25cm) -- (n5bl) arc (-90:-270:0.25cm);
\draw[fill=black!10] (n6tl) -- (n6tr) arc (90:-90:0.25cm) -- (n6bl) arc (-90:-270:0.25cm);

\vertex [above=0cm of n1b] {\(1\)};
\vertex [above=0cm of n2b] {\(N\)};
\vertex [right=0.5cm of n3bl] (n3b);
\vertex [right=0.5cm of n3tl] (n3t);
\vertex [above=0cm of n3b] {\(N+1\)};
\vertex [above=0cm of n4b] {\(1\)};
\vertex [above=0cm of n5b] {\(M\)};
\vertex [right=0.5cm of n6bl] (n6b);
\vertex [right=0.5cm of n6tl] (n6t);
\vertex [above=0cm of n6b] {\(M+1\)};

\diagram* {
  (tt) -- [very thick] (bb),
};
\end{feynman}
\end{tikzpicture}
\caption{The form, after using the Ward identities, of a general elastic scattering in the amplitude and in the complex conjugate amplitude. \label{fig:11}}
\end{figure}
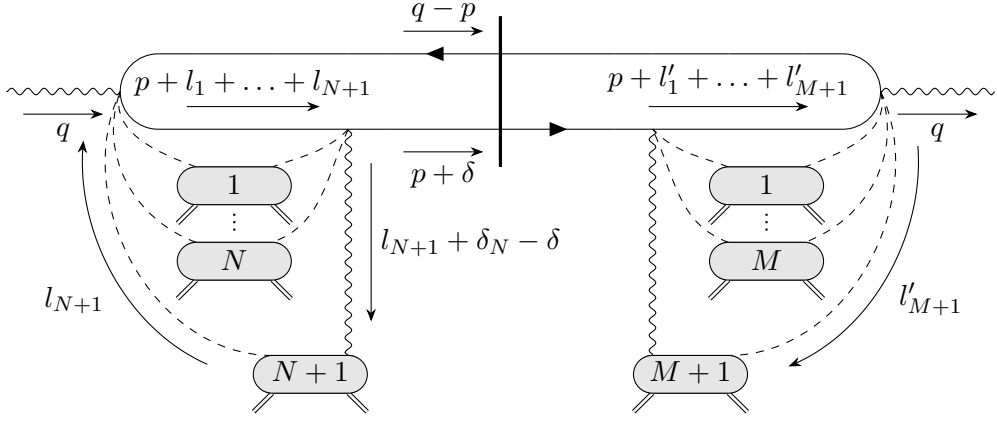

It should now be clear that for an arbitrary scattering the general coherent scattering graph has the form shown in Fig.~\ref{fig:11} where there are $N+1$ scatterings in the amplitude and $M+1$ in the complex conjugate amplitude.

Comparing the graphs in Fig.~\ref{fig:11} with that of Fig.~\ref{fig:6B} one sees that the fermion loop is identical in the two cases if one makes the replacement
\begin{equation} \label{eq:40}
\begin{split}
    l &\to l_1 + l_2 + \ldots + l_{N+1} \\
    l' &\to l'_1 + l'_2 + \ldots + l'_{M+1} \,.
\end{split}
\end{equation}
Thus we can get the case of an arbitrary number of scatterings from \eqref{eq:29} simply by calling $l_{N+1}=l$, $l'_{M+1}=l'$ and inserting factors $F(l_i)$ and $F(l'_i)$ as well as integrating over $\dd[2]{l_i}\dd[2]{l'_j}$ and making the replacement \eqref{eq:40} in the exponentials. Thus
\begin{multline} \label{eq:41}
    A = \frac{1}{128\pi^4q_+p_+\delta_-^2} \sum_{N=0}^\infty \int \frac{\dd[2]{l_1}\ldots\dd[2]{l_N}}{(N+1)!} \sum_{M=0}^\infty \frac{\dd[2]{l'_1}\ldots\dd[2]{l'_M}}{(M+1)!} e^{i\underline{x} {\cdot} \qty(\underline{l}_1+\ldots+\underline{l}_N) - i\underline{x}' {\cdot} \qty(\underline{l}'_1+\ldots+\underline{l}'_M)} \\
    \cdot F(\underline{l}_1) \ldots F(\underline{l}_N) \, F(\underline{l}'_1) \ldots F(\underline{l}'_M) \int \dd[2]{l} \dd[2]{l'} \dd[2]{x} \dd[2]{x'} F(\underline{l}) F(\underline{l}') e^{i\qty(\underline{p}+\underline{l}) {\cdot} \underline{x} - i\qty(\underline{p}+\underline{l}') {\cdot} \underline{x}'} \tr{\ldots}
\end{multline}
where the $\tr{\ldots}$ in \eqref{eq:41} is identical to that in \eqref{eq:29}. The trace in \eqref{eq:41} has $\underline{l}$ and $\underline{l}'$ exactly as in \eqref{eq:29} but $\underline{l}_i$ and $\underline{l}'_j$ do not appear.

In order to get the gluon distribution of the nucleon we set
\begin{equation} \label{eq:42}
\begin{split}
    e^{i\underline{l}_i{\cdot}\underline{x}} &\to -\frac{1}{2} \qty(\underline{l}_i{\cdot}\underline{x})^2 \ \to -\frac{1}{4} \underline{l}_i^2 \underline{x}^2 \\
    e^{i\underline{l}'_j{\cdot}\underline{x}'} &\to -\frac{1}{2} \qty(\underline{l}'_j{\cdot}\underline{x}')^2 \to -\frac{1}{4} \underline{l}_j^{\prime 2} \underline{x}^{\prime 2} \,.
\end{split}
\end{equation}
Then the sums over $M$ and $N$ can be done, and using \eqref{eq:22} one can write $A$ in \eqref{eq:41} as
\begin{multline} \label{eq:43}
    A = \frac{1}{128\pi^4q_+p_+\delta_-^2} \int \frac{\qty(e^{-Q_S^2x_\perp^2/4}-1)}{Q_S^2x_\perp^2/4} \frac{\qty(e^{-Q_S^2x_\perp^{\prime 2}/4}-1)}{Q_S^2x_\perp^{\prime 2}/4} \dd[2]{x} \dd[2]{x'} \\
    \int \dd[2]{l} \dd[2]{l'} F(\underline{l}) F(\underline{l}') e^{i\qty(\underline{p}+\underline{l}) {\cdot} \underline{x} - i\qty(\underline{p}+\underline{l}') {\cdot} \underline{x}'} \tr{\ldots} \,.
\end{multline}
\eqref{eq:43} differs from \eqref{eq:29} only in the factor
\begin{equation} \label{eq:44}
    \frac{\qty(e^{-Q_S^2x_\perp^2/4}-1)}{Q_S^2x_\perp^2/4} \frac{\qty(e^{-Q_S^2x_\perp^{\prime 2}/4}-1)}{Q_S^2x_\perp^{\prime 2}/4}
\end{equation}
so that our final answer for $A$ can be obtained simply by inserting \eqref{eq:44} into \eqref{eq:23} and using the fact that the $T(x_\perp)$ in \eqref{eq:23} is given by \eqref{eq:24}. Now calling
\begin{equation} \label{eq:45}
    T(x_\perp) = 1 - e^{-Q_S^2x_\perp^2/4}
\end{equation}
for the multiple scattering case we get 
\begin{equation} \label{eq:46}
    A = \bar{Q}^2 \int \frac{\dd[2]{x}\dd[2]{x'}}{(2\pi)^4} e^{i\underline{p}{\cdot}\qty(\underline{x}-\underline{x'})} K_1\qty(\bar{Q}x_\perp) \frac{\underline{x}{\cdot}\underline{x}'}{x_\perp x'_\perp} K_1\qty(\bar{Q}x'_\perp) T(x_\perp) T(x'_\perp)
\end{equation}
exactly the same structure as \eqref{eq:23} but where now $T$ is given by \eqref{eq:45} rather than \eqref{eq:24}.

Finally we note agreement between \eqref{eq:3} and \eqref{eq:46}. It was easy to get \eqref{eq:3} and rather difficult to obtain \eqref{eq:46}. However, the partonic picture, including saturation, is hidden in the derivation giving \eqref{eq:3} while we shall shortly see that the derivation of \eqref{eq:46} makes the partonic picture and partonic saturation manifest.

\subsection{Inelastic single scattering}
\label{sec:3.4}

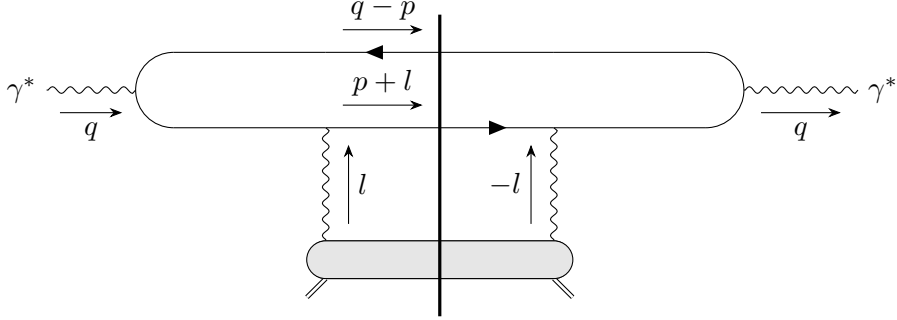
\begin{figure}[htbp]
\centering
\begin{tikzpicture}
\begin{feynman} 
\vertex (p1) {\(\gamma^*\)};
\vertex [right=of p1] (l);
\vertex [below right=0.707cm of l] (b1);
\vertex [right=2.0cm of b1] (b2);
\vertex [right=1.5cm of b2] (b3);
\vertex [right=1.5cm of b3] (b4);
\vertex [right=2.0cm of b4] (b5);
\vertex [above right=0.707cm of b5] (r);
\vertex [right=of r] (p2) {\(\gamma^*\)};
\vertex [above right=0.707cm of l] (t1);
\vertex [right=2.0cm of t1] (t2);
\vertex [right=1.5cm of t2] (t3);
\vertex [right=1.5cm of t3] (t4);
\vertex [right=2.0cm of t4] (t5);
\vertex [above=0.5cm of t3] (tt);
\vertex [below=2.5cm of b3] (bb);
\vertex [below=of b2] (nctl);
\vertex [below=of b4] (nctr);
\vertex [below=0.5cm of nctl] (ncbl);
\vertex [below=0.5cm of nctr] (ncbr);
\vertex [below left=0.35cm of ncbl] (ncl);
\vertex [below right=0.35cm of ncbr] (ncr);
\vertex [below=3.0cm of l] (brl);
\vertex [below=2.5cm of b5] (brr);

\diagram* {
  (l) -- [quarter right] (b1) -- (b2) -- [momentum=\(p+l\)] (b3) -- [fermion] (b4) -- (b5) -- [quarter right] (r),
  (l) -- [quarter left] (t1) -- (t2) -- [anti fermion, momentum=\(q-p\)] (t3) -- (t4) -- (t5) -- [quarter left] (r),
  (p1) -- [photon, momentum'=\(q\)] (l),
  (r) -- [photon, momentum'=\(q\)] (p2),
  (nctl) -- [photon, momentum'=\(l\)] (b2),
  (nctr) -- [photon, momentum=\(-l\)] (b4),
  (nctl) -- (nctr) -- [half left] (ncbr) -- (ncbl) -- [half left] (nctl),
  (ncbl) -- [double, double distance=1.0pt] (ncl),
  (ncbr) -- [double, double distance=1.0pt] (ncr),
};

\draw[fill=black!10] (nctl) -- (nctr) arc (90:-90:0.25cm) -- (ncbl) arc (-90:-270:0.25cm);

\diagram* {
  (tt) -- [very thick] (bb),
};
\end{feynman}
\end{tikzpicture}
\caption{An illustration of a single inelastic scattering. \label{fig:12}}
\end{figure}

The lowest order inelastic reaction is shown in Fig.~\ref{fig:12}. As usual we take $z$ zmall, to be in the logarithmic region of $p_\perp$, and in $A_-=0$ gauge the target only couples to the $p$-line and not to the $(q-p)$-line. We begin with the trace

\begin{equation} \label{eq:47}
    \tr = \tr{\gamma{\cdot}p\ \underline{\gamma}{\cdot}\underline{l}\ \gamma{\cdot}(p+l)\ \underline{\gamma}{\cdot}\underline{l}\ \gamma{\cdot}p\ \underline{\gamma}{\cdot}\underline{\epsilon}_\gamma\ \gamma{\cdot}(q-p)\ \underline{\gamma}{\cdot}\underline{\epsilon}_\gamma^*} \,.
\end{equation}
As usual $\gamma{\cdot}(q-p) \simeq \gamma_-q_+$ and $\underline{\gamma}{\cdot}\underline{\epsilon}_\gamma\ \gamma_-\ \underline{\gamma}{\cdot}\underline{\epsilon}_\gamma^* = \gamma_-$ so
\begin{equation} \label{eq:48}
    \tr = q_+ \tr{\gamma{\cdot}p\ \underline{\gamma}{\cdot}\underline{l}\ \gamma{\cdot}(p+l)\ \underline{\gamma}{\cdot}\underline{l}\ \gamma{\cdot}p\ \gamma_-} \,.
\end{equation}
In the end we shall average over the directions of $\underline{l}$ so
\begin{equation} \label{eq:49}
    \underline{\gamma}{\cdot}\underline{l}\ \gamma{\cdot}(p+l)\ \underline{\gamma}{\cdot}\underline{l} \to \underline{l}^2 \qty[\gamma_+(p+l)_- + \gamma_-p_+]
\end{equation}
and
\begin{equation*}
    \tr = q_+\underline{l}^2 \qty[(p+l)_-\tr{\gamma{\cdot}p\ \gamma_+\gamma{\cdot}p\ \gamma_-} + p_+\tr{\gamma{\cdot}p\ \gamma_-\gamma{\cdot}p\ \gamma_-}] \,.
\end{equation*}
Thus, using $(p+l)_l = \frac{\underline{p}^2}{2p_+}$ and $p_- \simeq -\frac{Q^2}{2q_+}$, one finds
\begin{equation} \label{eq:50}
    \tr \simeq \frac{2q_+\underline{l}^2}{p_+} \qty[\qty(\underline{p}^2)^2 + \qty(\bar{Q}^2)^2] \,.
\end{equation}
Also
\begin{equation*}
    \dd{p_-} \delta\qty[(q-p)^2] = \frac{1}{2q_+}
\end{equation*}
\begin{equation*}
    \dd{l_-} \delta\qty[(p+l)^2] = \frac{1}{2p_+}
\end{equation*}
so that
\begin{equation} \label{eq:51}
    p^2 = -\qty[\underline{p}^2 + \bar{Q}^2]
\end{equation}
and also
\begin{equation} \label{eq:52}
    l_- = \frac{1}{2p_+} \qty[\underline{p}^2 + \bar{Q}^2] \,.
\end{equation}
Then
\begin{equation} \label{eq:53}
    \frac{1}{2p_+2q_+} \frac{1}{l_-^2} \frac{\tr}{\qty[p^2]^2} = 2\underline{l}^2 \frac{\qty(\underline{p}^2)^2 + \qty(\bar{Q}^2)^2}{\qty[\underline{p}^2 + \bar{Q}^2]^4}
\end{equation}
and using
\begin{equation} \label{eq:53_}
    \frac{\qty(\underline{p}^2)^2 + \qty(\bar{Q}^2)^2}{\qty[\underline{p}^2 + \bar{Q}^2]^4} = \frac{\bar{Q}^2}{8\pi^2} \int \dd[2]{x_1} \dd[2]{x_2} e^{-i\underline{p}{\cdot}\qty(\underline{x}_1-\underline{x}_2)} K_1\qty(x_{1\perp}\bar{Q}) K_1\qty(x_{2\perp}\bar{Q}) \frac{\qty(\underline{x}_1{\cdot}\underline{x}_2)^2}{x_{1\perp}x_{2\perp}} \,,
\end{equation}
or 
\begin{equation} \label{eq:54}
    \int \dd[2]{p} \frac{\qty(\underline{p}^2)^2 + \qty(\bar{Q}^2)^2}{\qty[\underline{p}^2 + \bar{Q}^2]^4} = \frac{\bar{Q}^2}{2} \int \dd[2]{x} K_1^2\qty(x_\perp\bar{Q}) \underline{x}^2 \,,
\end{equation}
and including a factor $F(l_\perp)$ for the target nucleons, as well as using \eqref{eq:22}, one gets
\begin{equation} \label{eq:55}
    \frac{\dd{\sigma_{in}}}{\dd{z}\dd[2]{b}} = 2\alpha_{em}N_Ce_f^22\bar{Q}^2 \int \frac{\dd[2]{x}}{4\pi^2} K_1^2\qty(\bar{Q}x_\perp) \frac{Q_S^2 x_\perp^2}{4}
\end{equation}
where we have taken the limit $z \ll 1$. Eqs.~\eqref{eq:55} and \eqref{eq:5} are consistent if one drops the exponential in the last factor of \eqref{eq:5} and identifies $\int_0^L\dd{l} \hat{q} = \hat{q}L = Q_S^2$.

In our discussion of coherent reactions the transverse momentum of the forward jets has been fixed while in the inelastic case the jet transverse momentum has been integrated. In the coherent case there are no final state interactions so fixing the transverse momentum causes no essential complications. However, in the inelastic case the transverse momentum can be significantly modified by the final state interactions so that knowledge of the jet transverse momentum does not directly tell whether or not the inelastic scattering happened in the saturation regime.

\subsection{Inelastic reaction with multiple scattering}
\label{sec:3.5}

In \eqref{eq:5} the $e^{-2\hat{q}lx_\perp^2/4}$ factor represents elastic dipole-nuclear scattering, in the amplitude and complex conjugate amplitude, up to a distance $l$ in the nucleus. The derivation of \eqref{eq:5} in $A_+=0$ is straightforward. The object here is to reach a formula equivalent to \eqref{eq:5} in $A_-=0$ gauge. We begin with the graph of Fig.~\ref{fig:12} which we have just considered. Consider four classes of additional interactions, with another nucleon of the nucleus in addition to the one shown in Fig.~\ref{fig:12}. The two interactions of the $p$-line in Fig.~\ref{fig:12} with the additional nucleon can be: (i) Completely between the $l$ and $-l$ lines of Fig.~\ref{fig:12}. (ii) Before the $l$-line in the amplitude and between the $l$ and $-l$ lines in either the amplitude or complex conjugate amplitude. (iii) To the left of $l$ in the amplitude and to the right of $-l$ in the complex conjugate amplitude in Fig.~\ref{fig:12}. (iv) Either completely to the left of the line $l$ or completely to the right of line $(-l)$ in Fig.~\ref{fig:12}. The interactions of the additional nucleon with the $p$ line cancels in (i) and (ii) by adding the interactions between $l$ and $-l$ in the amplitude and complex conjugate amplitude. In (iii) the cancellation occurs by considering the additional interaction as the basic inelastic interaction and then adding the contribution of the $l$-lines as shown in Fig.~\ref{fig:12} to that where the $l$-line is moved to the right of the cut. Thus we are left only with additional interactions of type (iv), that is with additional elastic interactions in the amplitude and complex conjugate amplitude.

In $A_+=0$ gauge we could order the interactions sequentially as the (strongly Lorentz-contracted) dipole passed through the nucleus at rest. In $A_-=0$ gauge the interactions are not local in $x_+$ because of the $\frac{1}{l-i\epsilon}$ terms in the propagator, and with our choice of the $i\epsilon$ interactions of a nucleon with the dipole $(q-p,p)$ may occur long before the dipole has reached that nucleon \cite{7,8,9}. Thus we need another labelling for the nucleons distinct from the $l$ in the exponent in \eqref{eq:5}.

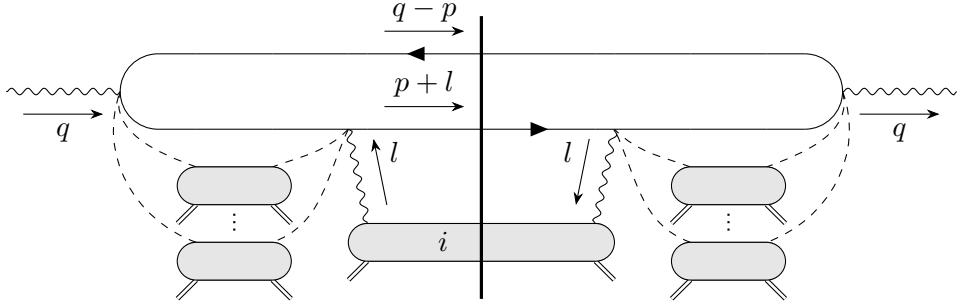
\begin{figure}[htbp]
\centering
\begin{tikzpicture}
\begin{feynman} 
\vertex (p1);
\vertex [right=of p1] (l);
\vertex [below right=0.707cm of l] (b1);
\vertex [right=0.5cm of b1] (b2);
\vertex [right=1.0cm of b2] (b3);
\vertex [right=1.0cm of b3] (b4);
\vertex [right=0.25cm of b4] (b5);
\vertex [right=1.5cm of b5] (b6);
\vertex [right=1.5cm of b6] (b7);
\vertex [right=0.25cm of b7] (b8);
\vertex [right=1.0cm of b8] (b9);
\vertex [right=1.0cm of b9] (b10);
\vertex [right=0.5cm of b10] (b11);
\vertex [above right=0.707cm of b11] (r);
\vertex [right=of r] (p2);
\vertex [above right=0.707cm of l] (t1);
\vertex [right=0.5cm of t1] (t2);
\vertex [right=1.0cm of t2] (t3);
\vertex [right=1.0cm of t3] (t4);
\vertex [right=0.25cm of t4] (t5);
\vertex [right=1.5cm of t5] (t6);
\vertex [right=1.5cm of t6] (t7);
\vertex [right=0.25cm of t7] (t8);
\vertex [right=1.0cm of t8] (t9);
\vertex [right=1.0cm of t9] (t10);
\vertex [right=0.5cm of t10] (t11);
\vertex [above=0.5cm of t6] (tt);
\vertex [below=2.25cm of b6] (bb);
\vertex [below=0.5cm of b2] (n1tl);
\vertex [below=0.5cm of b3] (n1tr);
\vertex [below=1.5cm of b2] (n2tl);
\vertex [below=1.5cm of b3] (n2tr);
\vertex [below=1.25cm of b5] (nctl);
\vertex [below=1.25cm of b7] (nctr);
\vertex [below=0.5cm of b9] (n4tl);
\vertex [below=0.5cm of b10] (n4tr);
\vertex [below=1.5cm of b9] (n5tl);
\vertex [below=1.5cm of b10] (n5tr);
\vertex [below=0.5cm of n1tl] (n1bl);
\vertex [below=0.5cm of n1tr] (n1br);
\vertex [below=0.5cm of n2tl] (n2bl);
\vertex [below=0.5cm of n2tr] (n2br);
\vertex [below=0.5cm of nctl] (ncbl);
\vertex [below=0.5cm of nctr] (ncbr);
\vertex [below=0.5cm of n4tl] (n4bl);
\vertex [below=0.5cm of n4tr] (n4br);
\vertex [below=0.5cm of n5tl] (n5bl);
\vertex [below=0.5cm of n5tr] (n5br);
\vertex [below left=0.35cm of n1bl] (n1l);
\vertex [below right=0.35cm of n1br] (n1r);
\vertex [below left=0.35cm of n2bl] (n2l);
\vertex [below right=0.35cm of n2br] (n2r);
\vertex [below left=0.35cm of ncbl] (ncl);
\vertex [below right=0.35cm of ncbr] (ncr);
\vertex [below left=0.35cm of n4bl] (n4l);
\vertex [below right=0.35cm of n4br] (n4r);
\vertex [below left=0.35cm of n5bl] (n5l);
\vertex [below right=0.35cm of n5br] (n5r);
\vertex [right=0.5cm of n1bl] (n1b);
\vertex [right=0.5cm of n1tl] (n1t);
\vertex [right=0.5cm of n2bl] (n2b);
\vertex [right=0.5cm of n2tl] (n2t);
\vertex [below=0.12cm of n1b] (n12t);
\vertex [above=0.12cm of n2t] (n12b);
\vertex [right=0.5cm of n4bl] (n4b);
\vertex [right=0.5cm of n4tl] (n4t);
\vertex [right=0.5cm of n5bl] (n5b);
\vertex [right=0.5cm of n5tl] (n5t);
\vertex [below=0.12cm of n4b] (n45t);
\vertex [above=0.12cm of n5t] (n45b);

\diagram* {
  (l) -- [quarter right] (b1) -- (b2) -- (b3) -- (b4) -- (b5) -- [momentum=\(p+l\)] (b6) -- [fermion] (b7) -- (b8) -- (b9) -- (b10) -- (b11) -- [quarter right] (r),
  (l) -- [quarter left] (t1) -- (t2) -- (t3) -- (t4) -- (t5) -- [anti fermion, momentum=\(q-p\)] (t6) -- (t7) -- (t8) -- (t9) -- (t10) -- (t11) -- [quarter left] (r),
  (p1) -- [photon, momentum'=\(q\)] (l),
  (r) -- [photon, momentum'=\(q\)] (p2),
  (n1tl) -- [scalar, out=160, in=-105] (l),
  (b4) -- [scalar, out=-135, in=20] (n1tr),
  (n2tl) -- [scalar, out=160, in=-110] (l),
  (b4) -- [scalar,out=-115, in=20] (n2tr),
  (nctl) -- [photon, momentum'=\(l\)] (b4),
  (b8) -- [photon, momentum'=\(l\)] (nctr),
  (n4tl) -- [scalar, out=160, in=-45] (b8),
  (r) -- [scalar, out=-75, in=20] (n4tr),
  (n5tl) -- [scalar, out=160, in=-65] (b8),
  (r) -- [scalar, out=-70, in=20] (n5tr),
  (n1tl) -- (n1tr) -- [half left] (n1br) -- (n1bl) -- [half left] (n1tl),
  (n2tl) -- (n2tr) -- [half left] (n2br) -- (n2bl) -- [half left] (n2tl),
  (nctl) -- (nctr) -- [half left] (ncbr) -- (ncbl) -- [half left] (nctl),
  (n4tl) -- (n4tr) -- [half left] (n4br) -- (n4bl) -- [half left] (n4tl),
  (n5tl) -- (n5tr) -- [half left] (n5br) -- (n5bl) -- [half left] (n5tl),
  (n1l) -- [double, double distance=1.0pt] (n1bl),
  (n1br) -- [double, double distance=1.0pt] (n1r),
  (n2l) -- [double, double distance=1.0pt] (n2bl),
  (n2br) -- [double, double distance=1.0pt] (n2r),
  (ncl) -- [double, double distance=1.0pt] (ncbl),
  (ncbr) -- [double, double distance=1.0pt] (ncr),
  (n4l) -- [double, double distance=1.0pt] (n4bl),
  (n4br) -- [double, double distance=1.0pt] (n4r),
  (n5l) -- [double, double distance=1.0pt] (n5bl),
  (n5br) -- [double, double distance=1.0pt] (n5r),
  (n12t) -- [ghost] (n12b),
  (n45t) -- [ghost] (n45b),
};

\draw[fill=black!10] (n1tl) -- (n1tr) arc (90:-90:0.25cm) -- (n1bl) arc (-90:-270:0.25cm);
\draw[fill=black!10] (n2tl) -- (n2tr) arc (90:-90:0.25cm) -- (n2bl) arc (-90:-270:0.25cm);
\draw[fill=black!10] (nctl) -- (nctr) arc (90:-90:0.25cm) -- (ncbl) arc (-90:-270:0.25cm);
\draw[fill=black!10] (n4tl) -- (n4tr) arc (90:-90:0.25cm) -- (n4bl) arc (-90:-270:0.25cm);
\draw[fill=black!10] (n5tl) -- (n5tr) arc (90:-90:0.25cm) -- (n5bl) arc (-90:-270:0.25cm);

\diagram* {
  (tt) -- [very thick] (bb),
};

\vertex [right=1.0cm of ncbl] (ncb);
\vertex [right=1.0cm of nctl] (nct);
\vertex [above=0cm of ncb] {\(i\)};
\end{feynman}
\end{tikzpicture}
\caption{The general graph contributing to the inelastic cross section. \label{fig:13}}
\end{figure}

Suppose, as the dipole passes through the nucleus it has the opportunity to interact with $N$ nucleons of the nucleus. Label these nucleons $1,2,\ldots N$ where the labelling is not necessarily related to the position that the nucleon occupies along the trajectory of the dipole $(q-p,p)$ as it passes through the nucleus. Then from the discussion we have just gone through we take only one of the nucleons to have an inelastic reaction. Call that nucleon $i$. Then if we allow nucleons $1,2,\ldots i-1$ either to interact elastically or not to interact at all we will have considered all possible interactions necessary to get the inelastic cross section by taking these interactions of nucleons $1,2,\ldots i-1$, along with the inelastic interaction of nucleon $i$, and summing over $i$ from $i=1$ to $i=N$. Thus we arrive at
\begin{equation} \label{eq:56}
    \frac{\dd{\sigma_{in}}}{\dd{z}\dd[2]{b}} = 2\alpha_{em}N_Ce_f^2\bar{Q}^2 \int \frac{\dd[2]{x}}{4\pi^2} K_1^2\qty(\bar{Q}x_\perp) \sum_{i=1}^N e^{-\frac{1}{2}\hat{q}\underline{x}^2 \qty(\frac{i-i}{N}L)} \frac{1}{2}\hat{q}\underline{x}^2 \frac{L}{N} \,.
\end{equation}
Calling $\frac{i}{N}L = l$ and $\dd{l} = \frac{L}{N}$ we get \eqref{eq:5}. The fact that the elastic interactions in the amplitude and complex amplitude arrange each to give a factor $e^{-\frac{1}{4}\hat{q}\qty(\frac{i}{N}L)}$ comes from evaluating these elastic interactions using the Ward identities, exactly as discussed in great detail in Sec.~\ref{sec:3.2} and Sec.~\ref{sec:3.3} above. Now the picture of the inelastic scattering evaluation in terms of graphs is given in Fig.~\ref{fig:13} where the $i^{\text{th}}$ nucleon scatters inelastically, as in \eqref{eq:56}, and nucleons $1\ldots (i-1)$ may scatter elastically, in the amplitude or complex conjugate amplitude, also as in \eqref{eq:56}.

\section{Target evolution in the $A_-=0$ gauge}
\label{sec:4}

We now choose a frame where the target nucleus momentum (per nucleon) $P_\mu$ is large
\begin{equation} \label{eq:57}
    P_\mu = \qty(\frac{m^2}{2P_-},P_-,0),\ \frac{m}{P_-} \ll 1
\end{equation}
while components of $q_\mu$ are, say, on the order of $Q$,
\begin{equation} \label{eq:58}
    q_\mu = \qty(q_+,-\frac{Q^2}{2q_+},0) \,.
\end{equation}
This is a natural frame to view a partonic picture of the scattering and the $A_-=0$ gauge is the gauge choice that makes the partonic picture of the target manifest. The momenta given in \eqref{eq:57} and \eqref{eq:58} are related to the momenta considered in the earlier sections of this note by a simple Lorentz boost along the $x_-$-axis and since the $A_-=0$ gauge is boost invariant the calculations necessary for our discussion have already been done in Sec.~\ref{sec:3}. The main change is that we now take $x_-$ to be the time variable while previously we had been considering $x_+$ to be the time variable. In the Feynman diagram calculations we have been doing this makes no difference. However, our $A_-=0$ calculations, viewed now with $x_-$ as a time variable, should agree with light cone perturbation theory calculations, with $A_-=0$ and $x_-$ the time variable, where the parton picture is manifest.

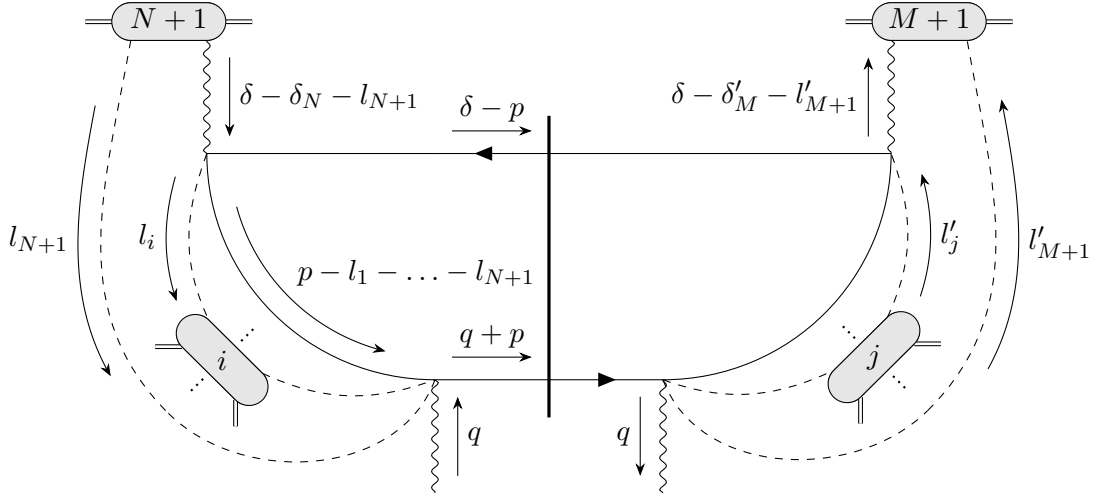
\begin{figure}[htbp]
\centering
\begin{tikzpicture}
\begin{feynman}
\vertex (l);
\vertex [right=3.0cm of l] (t1);
\vertex [right=1.5cm of t1] (t2);
\vertex [right=1.5cm of t2] (t3);
\vertex [right=3.0cm of t3] (r);
\vertex [below=3.0cm of t1] (b1);
\vertex [below=3.0cm of t2] (b2);
\vertex [below=3.0cm of t3] (b3);
\vertex [above=0.5cm of t2] (tt);
\vertex [below=0.5cm of b2] (bb);
\vertex [below=of b1] (p1);
\vertex [below=of b3] (p2);
\vertex [above=of l] (nNbr);
\vertex [left=1.0cm of nNbr] (nNbl);
\vertex [above=0.5cm of nNbl] (nNtl);
\vertex [above=0.5cm of nNbr] (nNtr);
\vertex [above left=0.35cm of nNbl] (nNl);
\vertex [left=0.35cm of nNl] (nNPl);
\vertex [above right=0.35cm of nNbr] (nNr);
\vertex [right=0.35cm of nNr] (nNPr);
\vertex [above= of r] (nMbl);
\vertex [right=1.0cm of nMbl] (nMbr);
\vertex [above=0.5cm of nMbl] (nMtl);
\vertex [above=0.5cm of nMbr] (nMtr);
\vertex [above left=0.35cm of nMbl] (nMl);
\vertex [left=0.35cm of nMl] (nMPl);
\vertex [above right=0.35cm of nMbr] (nMr);
\vertex [right=0.35cm of nMr] (nMPr);
\vertex [below=3.0cm of l] (nix);
\vertex [above right=0.0cm of nix] (nic) {\(i\)};
\vertex [above left=0.5cm of nic] (nil);
\vertex [above right=0.25cm of nil] (nitl);
\vertex [below left=0.25cm of nil] (nibl);
\vertex [below right=0.5cm of nic] (nir);
\vertex [above right=0.25cm of nir] (nitr);
\vertex [below left=0.25cm of nir] (nibr);
\vertex [left=0.35cm of nibl] (niPl);
\vertex [below=0.35cm of nibr] (niPr);
\vertex [above right=0.354cm of nic] (nit);
\vertex [above right=0.25cm of nit] (eit);
\vertex [below left=0.354cm of nic] (nib);
\vertex [below left=0.25cm of nib] (eib);
\vertex [below=3.0cm of r] (njx);
\vertex [above left=0.0cm of njx] (njc) {\(j\)};
\vertex [below left=0.5cm of njc] (njl);
\vertex [above left=0.25cm of njl] (njtl);
\vertex [below right=0.25cm of njl] (njbl);
\vertex [above right=0.5cm of njc] (njr);
\vertex [above left=0.25cm of njr] (njtr);
\vertex [below right=0.25cm of njr] (njbr);
\vertex [below=0.35cm of njbl] (njPl);
\vertex [right=0.35cm of njbr] (njPr);
\vertex [above left=0.354cm of njc] (njt);
\vertex [above left=0.25cm of njt] (ejt);
\vertex [below right=0.354cm of njc] (njb);
\vertex [below right=0.25cm of njb] (ejb);
\vertex [below left=1cm of nic] (nNp);
\vertex [below right=1cm of njc] (nMp);

\diagram* {
  (l) -- [quarter right, momentum=\(p-l_1-\ldots-l_{N+1}\)] (b1) -- [momentum=\(q+p\)] (b2) -- [fermion] (b3) -- [quarter right] (r),
  (l) -- (t1) -- [anti fermion, momentum=\(\delta-p\)] (t2) -- (t3) -- (r),
  (p1) -- [photon,momentum'=\(q\)] (b1),
  (b3) -- [photon,momentum'=\(q\)] (p2),
  (nNbr) -- [photon, momentum=\(\delta-\delta_N-l_{N+1}\)] (l),
  (r) -- [photon, momentum=\(\delta-\delta'_M-l'_{M+1}\)] (nMbl),
  (nNbl) -- [scalar, out=-100, in=135, momentum'=\(l_{N+1}\)] (nNp) -- [scalar, out=-45, in=-110,] (b1),
  (b3) -- [scalar, out=-70, in=-135] (nMp) -- [scalar, out=45, in=-80, momentum'=\(l'_{M+1}\)] (nMbr),
  (l) -- [scalar, out=-110, in=115, momentum'=\(l_i\)] (nitl),
  (nitr) -- [scalar, out=-25, in=-160] (b1),
  (b3) -- [scalar, out=-20, in=-155] (njtl),
  (njtr) -- [scalar, out=65, in=-70, momentum'=\(l'_j\)] (r),
  (nNtl) -- (nNtr) -- [half left] (nNbr) -- (nNbl) -- [half left] (nNtl),
  (nMtl) -- (nMtr) -- [half left] (nMbr) -- (nMbl) -- [half left] (nMtl),
  (nitl) -- (nitr) -- [half left] (nibr) -- (nibl) -- [half left] (nitl),
  (njtl) -- (njtr) -- [half left] (njbr) -- (njbl) -- [half left] (njtl),
  (nNl) -- [double, double distance=1.0pt] (nNPl),
  (nNr) -- [double, double distance=1.0pt] (nNPr),
  (nMl) -- [double, double distance=1.0pt] (nMPl),
  (nMr) -- [double, double distance=1.0pt] (nMPr),
  (nibl) -- [double, double distance=1.0pt] (niPl),
  (nibr) -- [double, double distance=1.0pt] (niPr),
  (njbl) -- [double, double distance=1.0pt] (njPl),
  (njbr) -- [double, double distance=1.0pt] (njPr),
  (nib) -- [ghost] (eib),
  (nit) -- [ghost] (eit),
  (njb) -- [ghost] (ejb),
  (njt) -- [ghost] (ejt),
};

\draw[fill=black!10] (nNtl) -- (nNtr) arc (90:-90:0.25cm) -- (nNbl) arc (-90:-270:0.25cm);
\draw[fill=black!10] (nMtl) -- (nMtr) arc (90:-90:0.25cm) -- (nMbl) arc (-90:-270:0.25cm);
\draw[fill=black!10] (nitl) -- (nitr) arc (45:-135:0.25cm) -- (nibl) arc (-135:-315:0.25cm);
\draw[fill=black!10] (njtl) -- (njtr) arc (135:-45:0.25cm) -- (njbl) arc (-45:-225:0.25cm);

\vertex [left=0.5cm of nNbr] (nNb);
\vertex [left=0.5cm of nNtr] (nNt);
\vertex [above=0.0cm of nNb] {\(N+1\)};
\vertex [left=0.5cm of nMbr] (nMb);
\vertex [left=0.5cm of nMtr] (nMt);
\vertex [above=0.0cm of nMb] {\(M+1\)};
\vertex [above right=0.0cm of nix] (nic) {\(i\)};
\vertex [above left=0.0cm of njx] (njc) {\(j\)};

\diagram* {
  (tt) -- [very thick] (bb),
};
\end{feynman}
\end{tikzpicture}
\caption{The same graph as shown in Fig.~\ref{fig:11} but now viewed as evolution of the target. \label{fig:14}}
\end{figure}

Thus Fig.~\ref{fig:11}, drawn with $x_+$ increasing from left to right in the amplitude, will now change to Fig.~\ref{fig:14} where we have changed $p\to-p$ in going from Fig.~\ref{fig:11} to Fig.~\ref{fig:14} to have $p_->0$ in Fig.~\ref{fig:14}. Now the process, in the amplitude, appears as if a nucleon, $N+1$, creates a quark-antiquark pair having $p_-$ and $(\delta_--p_-)$, and transverse momenta $\underline{p}$ and $-\underline{p}$, respectively. At its creation the quark also is given the transverse momentum $-\underline{l}_1-\underline{l}_2-\ldots-\underline{l}_N$ by the other $N$ nucleons which interact. The quark then propagates in $x_-$ until the $N$ nucleons give the quark $\underline{l}_1+\underline{l}_2+\ldots+\underline{l}_N$ bringing it back to $\underline{p}$ at which point it is struck by the $\gamma^*$ with \eqref{eq:45} and \eqref{eq:46} giving the precise formulas for the process as before.

Recalling that the photon scattering cross section and $F_2$ are related by
\begin{equation} \label{eq:59}
    \sigma_T^{\gamma^*} = \frac{4\pi^2\alpha_{em}}{Q^2} F_2\qty(x,Q^2)
\end{equation}
and using \eqref{eq:2} and \eqref{eq:3} one gets
\begin{equation} \label{eq:60}
    \frac{\dd{F_2^{f+\bar{f}}}}{\dd[2]{b}\dd[2]{p}} = \int_0^{1/2} \dd{z} \int \dd[2]{x}\dd[2]{x'} \frac{\bar{Q}^2Q^2N_Ce_f^2}{16\pi^6} e^{i\underline{p}{\cdot}\qty(\underline{x}-\underline{x}')} K_1\qty(\bar{Q}x_\perp) \frac{\underline{x}{\cdot}\underline{x}}{x_\perp x'_\perp} K_1\qty(\bar{Q}x'_\perp) T(x_\perp) T(x'_\perp)
\end{equation}
where we have included an extra factor of 2 in \eqref{eq:60} to account for measuring both quarks and antiquarks while \eqref{eq:2} refers only to a quark measurement. Take $p_\perp \ll Q_S$ in \eqref{eq:60} so that $T(x_\perp)$, $T(x'_\perp)$ are close to 1. Then writing $K_1\qty(\bar{Q}x_\perp) \frac{\underline{x}}{x_\perp} = \frac{1}{\bar{Q}} \nabla_{\underline{x}} K_0\qty(\bar{Q}x_\perp)$, and similarly for $K_1\qty(\bar{Q}x'_\perp)$, one can integrate $\nabla_{\underline{x}}$ and $\nabla_{\underline{x}'}$ by parts after which the $\dd[2]{x}$ and $\dd[2]{x'}$ integrals can be done. One gets
\begin{equation} \label{eq:61}
    \frac{\dd{F_2^{f+\bar{f}}}}{\dd[2]{b}\dd[2]{p}} = \frac{N_Ce_f^2}{4\pi^4} \,.
\end{equation}
Also
\begin{equation} \label{eq:62}
    F_2\qty(x,Q^2) = \sum_f e_f^2 \qty(xq_f\qty(x,Q^2)+x\bar{q}_f\qty(x,Q^2))
\end{equation}
so \eqref{eq:61} gives
\begin{equation} \label{eq:63}
    \frac{\dd}{\dd[2]{b}\dd[2]{p}} \qty(xq_f\qty(x,Q^2)+x\bar{q}_f\qty(x,Q^2))^{elastic} = \frac{N_C}{4\pi^4} \,.
\end{equation}

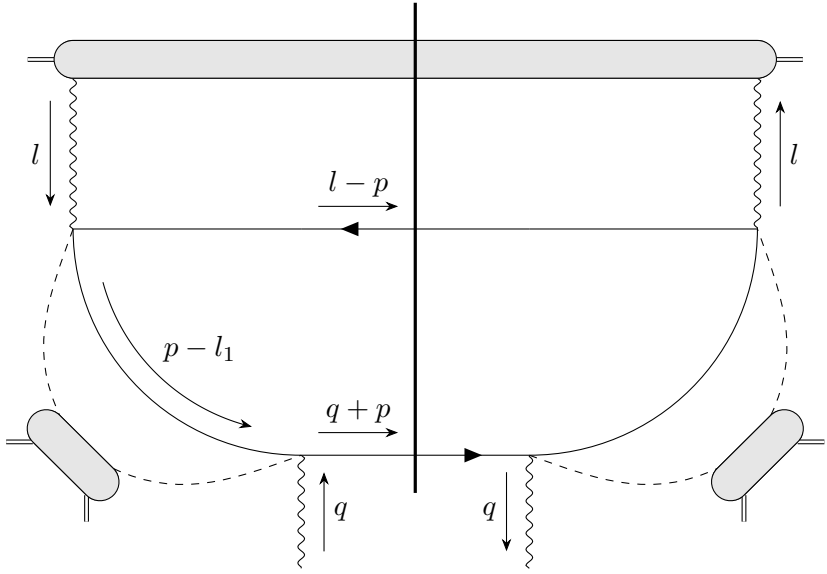
\begin{figure}[htbp]
\centering
\begin{tikzpicture}
\begin{feynman}
\vertex (l);
\vertex [right=3.0cm of l] (t1);
\vertex [right=1.5cm of t1] (t2);
\vertex [right=1.5cm of t2] (t3);
\vertex [right=3.0cm of t3] (r);
\vertex [below=3.0cm of t1] (b1);
\vertex [below=3.0cm of t2] (b2);
\vertex [below=3.0cm of t3] (b3);
\vertex [above=3.0cm of t2] (tt);
\vertex [below=0.5cm of b2] (bb);
\vertex [below=of b1] (p1);
\vertex [below=of b3] (p2);
\vertex [above=2.0cm of l] (ncbl);
\vertex [above=2.0cm of r] (ncbr);
\vertex [above=0.5cm of ncbl] (nctl);
\vertex [above=0.5cm of ncbr] (nctr);
\vertex [above left=0.35cm of ncbl] (ncl);
\vertex [left=0.35cm of ncl] (ncPl);
\vertex [above right=0.35cm of ncbr] (ncr);
\vertex [right=0.35cm of ncr] (ncPr);
\vertex [below=3.0cm of l] (nix);
\vertex [above right=0.0cm of nix] (nic);
\vertex [above left=0.5cm of nic] (nil);
\vertex [above right=0.25cm of nil] (nitl);
\vertex [below left=0.25cm of nil] (nibl);
\vertex [below right=0.5cm of nic] (nir);
\vertex [above right=0.25cm of nir] (nitr);
\vertex [below left=0.25cm of nir] (nibr);
\vertex [left=0.35cm of nibl] (niPl);
\vertex [below=0.35cm of nibr] (niPr);
\vertex [below=3.0cm of r] (njx);
\vertex [above left=0.0cm of njx] (njc);
\vertex [below left=0.5cm of njc] (njl);
\vertex [above left=0.25cm of njl] (njtl);
\vertex [below right=0.25cm of njl] (njbl);
\vertex [above right=0.5cm of njc] (njr);
\vertex [above left=0.25cm of njr] (njtr);
\vertex [below right=0.25cm of njr] (njbr);
\vertex [below=0.35cm of njbl] (njPl);
\vertex [right=0.35cm of njbr] (njPr);
\vertex [below left=1cm of nic] (nNp);
\vertex [below right=1cm of njc] (nMp);
\diagram* {
  (l) -- [quarter right, momentum=\(p-l_1\)] (b1) -- [momentum=\(q+p\)] (b2) -- [fermion] (b3) -- [quarter right] (r),
  (l) -- (t1) -- [anti fermion, momentum=\(l-p\)] (t2) -- (t3) -- (r),
  (p1) -- [photon,momentum'=\(q\)] (b1),
  (b3) -- [photon,momentum'=\(q\)] (p2),
  (ncbl) -- [photon, momentum'=\(l\)] (l),
  (r) -- [photon, momentum'=\(l\)] (ncbr),
  (l) -- [scalar, out=-110, in=115] (nitl),
  (nitr) -- [scalar, out=-25, in=-160] (b1),
  (b3) -- [scalar, out=-20, in=-155] (njtl),
  (njtr) -- [scalar, out=65, in=-70] (r),
  (nctl) -- (nctr) -- [half left] (ncbr) -- (ncbl) -- [half left] (nctl),
  (nitl) -- (nitr) -- [half left] (nibr) -- (nibl) -- [half left] (nitl),
  (njtl) -- (njtr) -- [half left] (njbr) -- (njbl) -- [half left] (njtl),
  (ncl) -- [double, double distance=1.0pt] (ncPl),
  (ncr) -- [double, double distance=1.0pt] (ncPr),
  (nibl) -- [double, double distance=1.0pt] (niPl),
  (nibr) -- [double, double distance=1.0pt] (niPr),
  (njbl) -- [double, double distance=1.0pt] (njPl),
  (njbr) -- [double, double distance=1.0pt] (njPr),
};

\draw[fill=black!10] (nitl) -- (nitr) arc (45:-135:0.25cm) -- (nibl) arc (-135:-315:0.25cm);
\draw[fill=black!10] (njtl) -- (njtr) arc (135:-45:0.25cm) -- (njbl) arc (-45:-225:0.25cm);
\draw[fill=black!10] (nctl) -- (nctr) arc (90:-90:0.25cm) -- (ncbl) arc (-90:-270:0.25cm);

\diagram* {
  (tt) -- [very thick] (bb),
};
\end{feynman}
\end{tikzpicture}
\caption{An illustration of the inelastic cross section. \label{fig:15}}
\end{figure}

It should be emphasized that \eqref{eq:63} corresponds to the quark and antiquark phase space distributions for coherent reactions. One can do a similar calculation for inelastic reactions, illustrated in Fig.~\ref{fig:15}. In the formalism we have used only one inelastic interaction with a nucleon is included. This is sufficient to generate the inelastic cross section, or $F_2^{inel}$, but not to give the distribution of the final jets in transverse momentum. However, it should be sufficient to generate the momentum distribution of the struck quark and it is to that calculation that we now turn.

We start with \eqref{eq:5}, which is the same is \eqref{eq:56}. Change the $K_1\qty(\bar{Q}x_\perp)$ in \eqref{eq:5} to $K_0$ by
\begin{equation} \label{eq:64}
    K_1^2\qty(\bar{Q}x_\perp) = \frac{1}{\bar{Q}^2} \nabla_{\underline{x}} K_0\qty(\bar{Q}x_\perp) \cdot \nabla_{\underline{x}} K_0\qty(\bar{Q}x_\perp)
\end{equation}
and write
\begin{equation} \label{eq:65}
    K_0\qty(\bar{Q}x_\perp) = \int \frac{\dd[2]{p}}{2\pi} \frac{e^{i\underline{p}{\cdot}\underline{x}}}{\underline{p}^2+\bar{Q}^2}
\end{equation}
to get \eqref{eq:5} in the form
\begin{equation} \label{eq:66}
    \dv[2]{\sigma_{in}}{b} = 2\alpha_{em}N_Ce_f^2\bar{Q}^2 \int_0^{1/2} 2 \dd{z} \int \frac{\dd[2]{x}}{4\pi^2} \int \frac{\dd[2]{p}\dd[2]{p'}}{4\pi^2\bar{Q}^2} \frac{e^{i\qty(\underline{p}-\underline{p}'){\cdot}\underline{x}}}{\qty(\underline{p}^2+\bar{Q}^2) \qty(\underline{p}^{\prime 2}+\bar{Q}^2)} \,.
\end{equation}
It will be explained shortly why the exponential in \eqref{eq:5} has been dropped. The integral over $z$ has been restricted to $0<z<\frac{1}{2}$ and a factor 2 inserted to cover the region $\frac{1}{2} < z < 1$. The $\dd[2]{x}$ integration now can be done trivially giving
\begin{equation} \label{eq:67}
    \frac{\dd{\sigma_{in}^{\gamma^*}}}{\dd[2]{b}\dd[2]{p}} = \frac{\alpha_{em}}{\pi^2} N_Ce_f^2 \int_0^{1/2} \dd{z} \frac{\underline{p}^2}{\qty(\underline{p}^2+Q^2z)^2}
\end{equation}
or
\begin{equation} \label{eq:68}
    \frac{\dd{\sigma_{in}^{\gamma^*}}}{\dd[2]{b}\dd[2]{p}} = \frac{\alpha_{em}N_Ce_f^2}{\pi^2Q^2} \,.
\end{equation}
Using \eqref{eq:59}
\begin{equation} \label{eq:69}
    \frac{\dd{F_2^{f+\bar{f}}}}{\dd[2]{b}\dd[2]{p}} = \frac{N_Ce_f^2}{4\pi^4}
\end{equation}
or
\begin{equation} \label{eq:70}
    \frac{\dd}{\dd[2]{b}\dd[2]{p}} \qty(xq_f+x\bar{q}_f)^{inelastic} = \frac{N_C}{4\pi^4}
\end{equation}
exactly the same is \eqref{eq:63}. \eqref{eq:63} refers to the distribution of quarks in the wave function which are part of a zero total momentum and zero color charge quark-antiquark pair in the light cone wave function while \eqref{eq:70} refers to quark, and antiquarks, which are not part of such zero momentum and zero charge pairs. Now back to why the final term in \eqref{eq:5} was dropped in \eqref{eq:66}. We are looking at a region where $\underline{p}^2 < Q_S^2$ in \eqref{eq:66} and that means that $x_\perp$ in \eqref{eq:66}, and in \eqref{eq:5} is large. The $\dd{l}$ integral in \eqref{eq:5} is then only evaluated at the lower limit, at $l=0$, causing the whole integral in the last term in \eqref{eq:5} to be unity.

Finally we note that if \eqref{eq:63} and \eqref{eq:70} are added to give the full quark, and antiquark, distributions in the nucleus the result is the same as (29) of \cite{2} where there was no separation made between elastic and inelastic events.

\section{The phenomenon of saturation}
\label{sec:5}

In this section an attempt will be made to interpret the results and picture of the previous sections. How to go beyond our MV model is a topic which will also be discussed.

\subsection{Coherent states}
\label{sec:5.1}

In our earlier discussion we have often talked about coherent scattering where the nucleus remains in its ground state after the scattering but (at least) two jets are produced along the $\gamma^*$ direction. Refer now to the coherent scattering process in Fig.~\ref{fig:14} as viewed in a Bjorken-type frame. The nucleons $i=1$ to $i=N+1$ are part of a large nucleus which does not break up after the scattering. Viewed in terms of the light cone wave function of the nucleus the quark-antiquark pair of Fig.~\ref{fig:14} is part of the nuclear wave function just before the $\gamma^*$ strikes the quark. The $\gamma^*$ frees the quark-antiquark pair but does not break up the nucleus. This can occur because the quark-antiquark pair, just before the scattering, is in a color singlet state with total transverse momentum almost zero. One might have guessed that such zero color, zero transverse momentum pairs would be rare but our calculation of their probability shows that that is not the case. Indeed, comparing \eqref{eq:63} and \eqref{eq:70} one sees that one-half of all quarks, having transverse momentum less than the saturation momentum are part of such zero color, zero transverse momentum pairs. Thus at small $x$ and for quark transverse momentum less than the saturation momentum there are many quark-antiquark pairs having zero total color and transverse momentum in the nuclear wave function which are so coherently arranged that if a photon frees one of these pairs the coherence of the overall nuclear wave function is not broken and the nucleus does not break up.

Recall a simple example of a coherent state. If $a$ and $a^\dagger$ are an annihilation and a creation operator, respectively, then
\begin{equation} \label{eq:71}
    \ket{S} = e^{\lambda a^\dagger} \ket{0}
\end{equation}
is a coherent state, that is
\begin{equation} \label{eq:72}
    a\ket{S} = \lambda\ket{S} \,.
\end{equation}
For our purposes is is better to write \eqref{eq:72} in terms of the number operator $N=a^\dagger a$, then
\begin{equation} \label{eq:73}
    N\ket{S} = \lambda a^\dagger \ket{S} \,.
\end{equation}
\eqref{eq:73} is exactly what is happening in our $\gamma^*$-nucleus process. The $\gamma^*$ is like a number operator acting on the nucleus which frees a $(q\bar{q})$ pair, the $a^\dagger$ in \eqref{eq:73}, while leaving the nucleus, the state $\ket{S}$ in \eqref{eq:73}, unchanged. It would appear that a dipole scattering in the black disc limit is dual to the nucleus having a coherent state of zero momentum dipoles as part of its wave function.

\subsection{Recombinations}
\label{sec:5.2}

\begin{figure}[htbp]
\centering
\begin{tikzpicture}
\begin{feynman}
\vertex (l);
\vertex [right=6.0cm of l] (t1);
\vertex [right=of t1] (tr);
\vertex [below right=2.121cm of l] (b1);
\vertex [below right=2.121cm of b1] (b2);
\vertex [below right=2.121cm of b2] (b3);
\vertex [below right=2.121cm of b3] (b4);
\vertex [left=2.0cm of b2] (bx);
\vertex [right=of b4] (br);
\vertex [below left=of b4] (p);
\vertex [below right=0.5cm of l] (hb);
\vertex [above left=1.5cm of l] (ht);
\vertex [below left=2.0cm of l] (n2x);
\vertex [above left=2.121cm of l] (n2br);
\vertex [left=1.0cm of n2br] (n2bl);
\vertex [above=0.5cm of n2bl] (n2tl);
\vertex [above=0.5cm of n2br] (n2tr);
\vertex [above left=0.35cm of n2bl] (n2l);
\vertex [left=0.35cm of n2l] (n2Pl) {\(P_2\)};
\vertex [above right=0.35cm of n2br] (n2r);
\vertex [right=0.35cm of n2r] (n2Pr) {\(P_2-\delta+\delta_1\)};
\vertex [below left=2.0cm of b2] (n1x);
\vertex [above right=0.0cm of n1x] (n1c);
\vertex [above left=0.5cm of n1c] (n1l);
\vertex [above right=0.25cm of n1l] (n1tl);
\vertex [below left=0.25cm of n1l] (n1bl);
\vertex [below right=0.5cm of n1c] (n1r);
\vertex [above right=0.25cm of n1r] (n1tr);
\vertex [below left=0.25cm of n1r] (n1br);
\vertex [left=0.35cm of n1bl] (n1Pl) {\(P_1\)};
\vertex [below=0.35cm of n1br] (n1Pr) {\(P_1-\delta_1\)};

\diagram* {
  (l) -- [anti fermion] (t1) -- [momentum=\(\delta-p\)] (tr),
  (l) -- (hb) -- [momentum=\(p-\delta_1-l_2\)] (b1) -- [momentum=\(p-l_1-l_2\)] (b2) -- [fermion,momentum=\(p-l_1\)] (b3) -- [momentum=\(p\)] (b4) -- [momentum'=\(q+p\)] (br),
  (p) -- [photon,momentum=\(q\)] (b4),
  (l) -- [photon,momentum'=\(l_2+\delta_1-\delta\)] (ht) -- [photon] (n2br),
  (n2bl) -- [photon, out=-100, in=130, momentum'=\(l_2\)] (n2x) -- [photon, out=-50, in=-160] (b2),
  (b1) -- [photon, out=-135, in=90, momentum'=\(l_1-\delta_1\)] (bx) -- [photon, out=-90, in=115] (n1tl),
  (n1tr) -- [photon, out=-25, in=-135, momentum'=\(l_1\)] (b3),
  (n1tl) -- (n1tr) -- [half left] (n1br) -- (n1bl) -- [half left] (n1tl),
  (n2tl) -- (n2tr) -- [half left] (n2br) -- (n2bl) -- [half left] (n2tl),
  (n1bl) -- [double, double distance=1.0pt] (n1Pl),
  (n1br) -- [double, double distance=1.0pt] (n1Pr),
  (n2l) -- [double, double distance=1.0pt] (n2Pl),
  (n2r) -- [double, double distance=1.0pt] (n2Pr),
};

\draw[fill=black!10] (n1tl) -- (n1tr) arc (45:-135:0.25cm) -- (n1bl) arc (-135:-315:0.25cm);
\draw[fill=black!10] (n2tl) -- (n2tr) arc (90:-90:0.25cm) -- (n2bl) arc (-90:-270:0.25cm);

\vertex [left=0.5cm of n2br] (n2b);
\vertex [left=0.5cm of n2tr] (n2t);
\vertex [above=0.0cm of n2b] {\(2\)};
\vertex [below right=0.15cm of n1l] (n1c) {\(1\)};
\end{feynman}
\end{tikzpicture}
\caption{The same graph as shown in Fig.~\ref{fig:7B} but now viewed in terms of target time rather than projectile time. \label{fig:16}}
\end{figure}
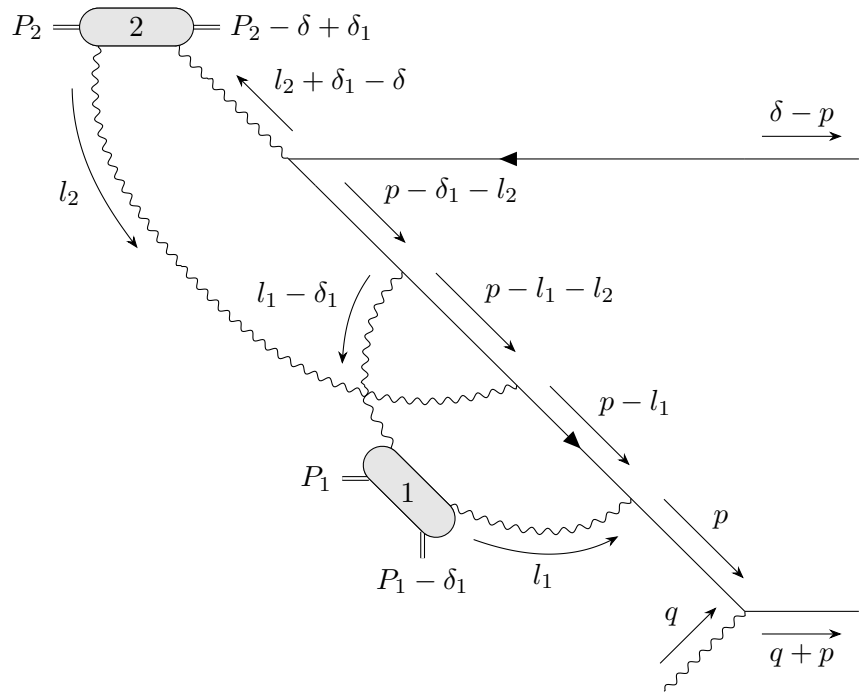

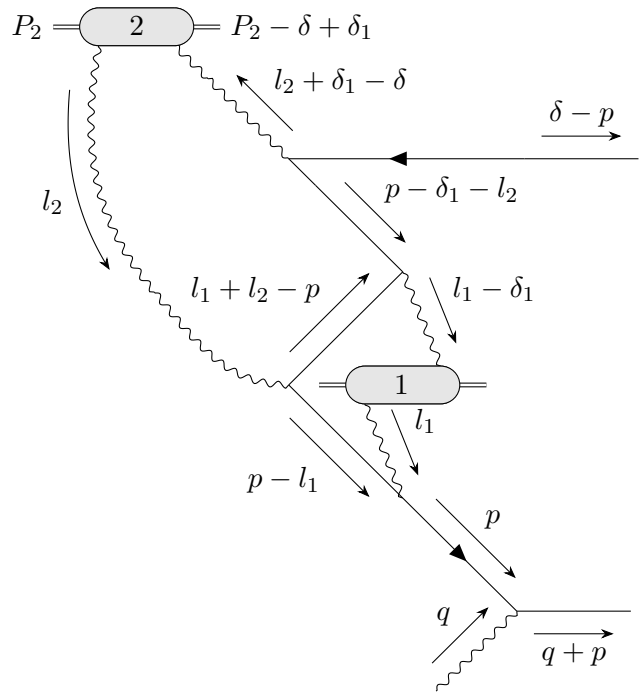
\begin{figure}[htbp]
\centering
\begin{tikzpicture}
\begin{feynman}
\vertex (l);
\vertex [right=3.1cm of l] (t1);
\vertex [right=of t1] (tr);
\vertex [below right=2.121cm of l] (b1);
\vertex [below left=2.121cm of b1] (b2);
\vertex [below right=2.121cm of b2] (b3);
\vertex [below right=2.121cm of b3] (b4);
\vertex [left=2.0cm of b2] (bx);
\vertex [right=of b4] (br);
\vertex [below left=of b4] (p);
\vertex [below right=0.5cm of l] (hb);
\vertex [above left=1.5cm of l] (ht);
\vertex [below left=2.5cm of l] (n2x);
\vertex [above left=2.121cm of l] (n2br);
\vertex [left=1.0cm of n2br] (n2bl);
\vertex [above=0.5cm of n2bl] (n2tl);
\vertex [above=0.5cm of n2br] (n2tr);
\vertex [above left=0.35cm of n2bl] (n2l);
\vertex [left=0.35cm of n2l] (n2Pl) {\(P_2\)};
\vertex [above right=0.35cm of n2br] (n2r);
\vertex [right=0.35cm of n2r] (n2Pr) {\(P_2-\delta+\delta_1\)};
\vertex [below=1.5cm of b1] (n1x);
\vertex [above right=0.0cm of n1x] (n1c);
\vertex [left=0.5cm of n1c] (n1l);
\vertex [above=0.25cm of n1l] (n1tl);
\vertex [below=0.25cm of n1l] (n1bl);
\vertex [right=0.5cm of n1c] (n1r);
\vertex [above=0.25cm of n1r] (n1tr);
\vertex [below=0.25cm of n1r] (n1br);
\vertex [left=0.6cm of n1l] (n1Pl);
\vertex [right=0.6cm of n1r] (n1Pr);

\diagram* {
  (l) -- [anti fermion] (t1) -- [momentum=\(\delta-p\)] (tr),
  (l) -- (hb) -- [momentum=\(p-\delta_1-l_2\)] (b1),
  (b2) -- [momentum=\(l_1+l_2-p\)] (b1),
  (b2) -- [momentum'=\(p-l_1\)] (b3) -- [fermion, momentum=\(p\)] (b4) -- [momentum'=\(q+p\)] (br),
  (p) -- [photon,momentum=\(q\)] (b4),
  (l) -- [photon,momentum'=\(l_2+\delta_1-\delta\)] (ht) -- [photon] (n2br),
  (n2bl) -- [photon, out=-100, in=130, momentum'=\(l_2\)] (n2x) -- [photon, out=-50, in=180] (b2),
  (b1) -- [photon, momentum=\(l_1-\delta_1\)] (n1tr),
  (n1bl) -- [photon, momentum=\(l_1\)] (b3),
  (n1tl) -- (n1tr) -- [half left] (n1br) -- (n1bl) -- [half left] (n1tl),
  (n2tl) -- (n2tr) -- [half left] (n2br) -- (n2bl) -- [half left] (n2tl),
  (n1l) -- [double, double distance=1.0pt] (n1Pl),
  (n1r) -- [double, double distance=1.0pt] (n1Pr),
  (n2l) -- [double, double distance=1.0pt] (n2Pl),
  (n2r) -- [double, double distance=1.0pt] (n2Pr),
};

\draw[fill=black!10] (n1tl) -- (n1tr) arc (90:-90:0.25cm) -- (n1bl) arc (-90:-270:0.25cm);
\draw[fill=black!10] (n2tl) -- (n2tr) arc (90:-90:0.25cm) -- (n2bl) arc (-90:-270:0.25cm);

\vertex [left=0.5cm of n2br] (n2b);
\vertex [left=0.5cm of n2tr] (n2t);
\vertex [above=0.0cm of n2b] {\(2\)};
\vertex [right=0.25cm of n1l] (n1c) {\(1\)};
\end{feynman}
\end{tikzpicture}
\caption{The same graph as in Fig.~\ref{fig:16} but now viewed as light cone perturbation theory graphs as opposed to Feynman graphs. \label{fig:17}}
\end{figure}

For many years the question has arisen as to whether parton saturation comes about because of partonic recombination when the parton densities become large\cite{10}. Our calculations do not give a definitive answer to that question but suggest strongly that recombination is a key element in saturation. Let's see how the above calculations have touched on this issue and what additional calculations would have to be done to give a definitive answer as to the importance of recombination. Refer back to Sec.~\ref{sec:3.3} where the graphs in Fig.~\ref{fig:7} were considered. We evaluated the double scattering graphs in Fig.~\ref{fig:7} by applying the Ward identities and reducing these graphs to the graph in Fig.~\ref{fig:8}. Let's focus on the graph of Fig.~\ref{fig:7B} which is redrawn in Fig.~\ref{fig:16} with $x_-$ being the time variable rather than having $x_+$ the time variable as in the graph of Fig.~\ref{fig:7B}, and we have let $p\to-p$ as always when going to the target evolution picture. Earlier when evaluating the graph of Fig.~\ref{fig:16} (or Fig.~\ref{fig:7B}) we used the Ward identities to reduce that graph to that of Fig.~\ref{fig:8}. Note that in Fig.~\ref{fig:8} there is only one quark propagator, $p-l_1-l_2$ in the target frame notation, while the graph in Fig.~\ref{fig:16} has four quark lines $p-\delta_1-l_2$, $p-l_1-l_2$, $p-l_1$ and $p$. If the minus component of any of these momenta is negative then the line becomes backward moving. This is illustrated in Fig.~\ref{fig:17} where the lines are now to be understood as lines in light cone perturbation theory and where the arrows on the lines follow the flow of the minus component of the momentum with the minus component, as written on each of the quark lines being positive. The graphs in Fig.~\ref{fig:16} and Fig.~\ref{fig:17} are equivalent. Fig.~\ref{fig:16} is a picture of a Feynman graph while Fig.~\ref{fig:17} is of a light cone perturbation theory graph. In addition to the graph shown in Fig.~\ref{fig:17} other graphs in light cone perturbation theory must be included to agree with the Feynman graph of Fig.~\ref{fig:16}. In Fig.~\ref{fig:17} at the vertex having a gluon $(l_1-\delta_1)$ a quark and an antiquark annihilate. This is a recombination graph. In Fig.~\ref{fig:17} nucleon 2 emits two separate pairs giving 2 quarks and 2 antiquarks. However, a quark and an antiquark annihilate into a gluon attached to nucleon 1 thus lowering the quark and antiquark number to a single quark-antiquark pair.

Because the minus component of the gluons in Fig.~\ref{fig:8} is zero, after having done contour distortions and used Ward identities on graphs such as that in Fig.~\ref{fig:16} we are unable to determine how important the annihilation graphs, as in Fig.~\ref{fig:17}, are. On the other hand since the minus component of the momenta on the quark and antiquark lines in Fig.~\ref{fig:16} and Fig.~\ref{fig:17} is small compare to that which can be given up or absorbed by the nucleons bound in a large nucleus, it is natural that light cone perturbation theory graphs, like that of Fig.~\ref{fig:17} should have multiple virtual creation and annihilation of pairs in each event. To have a good understanding of the role of recombination in saturation would require a direct evaluation of the light cone perturbation theory graphs from which the graphs of Fig.~\ref{fig:14} come about. This appears far beyond our current technical capabilities.

\subsection{Beyond the McLerran-Venugopalan model}
\label{sec:5.3}

The great virtue of the MV model is that we may imagine many nucleons, each well separated from the other along the $z$-direction (the projectile direction of motion) so that the scattering of one nucleon may be weak but, if we take a sufficient number of nucleons, the scattering of the projectile with all the nucleons at a given impact parameter is strong. The calculations outlined in Sec.~\ref{sec:2}, in the $A_+=0$ gauge, made use of this where the projectile, a quark-antiquark dipole coming from the $\gamma^*$, sequentially has the possibility of scattering on the various individual nucleons and because the scattering is sequential the formulas for the scattering are just exponentials of a single dipole nucleon scattering in a transverse coordinate basis for the dipole. However, when one uses $A_-=0$ gauge the idea of sequential scattering is lost since the scattering is very nonlocal in $x_+$ because of the $\frac{1}{l_-}$ terms in the gluon propagator. We have chosen the $i\epsilon$ that goes along with the $l_-$ so that final state interactions of the dipole with the nucleons are suppressed\cite{7,8,9}, but the dipole can interact with a nucleon long before the dipole reaches the target. This is apparent, for example, in Fig.~\ref{fig:8} where the interactions of the $l_1$ and $l_2$ lines reach the point where the dipole is created from the $\gamma^*$, long before reaching the target.

Also when we evaluated inelastic scattering in Sec.~\ref{sec:3.5} we labelled the various nucleons in the nucleus not necessarily by their position along the projectile dipole's path but by a more general labelling which distinguishes the different nucleons in the nucleus.

Since the discussion in Secs.~\ref{sec:3.3} and \ref{sec:3.5} did not use the details of the longitudinal positions of the nucleons on which the projectile dipole scatters one should be able to replace the MV nucleus by a collection of elementary dipoles which have arisen through dipole splitting in the target evolution of the nucleus. The problem with trying to do this technically and in detail is that unitarity effects on the scattering dipole will only arise when the density of dipoles in the nucleus is also in the saturation region. Thus one must deal with target evolution in the presence of saturation at exactly the same time that unitarity effects (saturation) occur in the scattering. The MV model avoids this by taking a target which is unevolved but which is big enough to create a saturation momentum $Q_S$ and where the individual nucleons are well separated so that they do not interact.

\acknowledgments

I have greatly benefited from conversations with Edmond Iancu and Dionysios Triantafyllopoulos on many issues related to this paper. I had very useful conversations with Bowen Xiao on an early draft of this paper. This work was supported in part by DOE grant DE-SC0011941.



\bibliographystyle{JHEP}
\bibliography{biblio.bib}






\end{document}